\documentclass[english,aps,manuscript,aps,preprint,amsmath]{revtex4}
\usepackage[T1]{fontenc}
\usepackage[latin9]{inputenc}
\usepackage{amsmath}
\usepackage{amssymb}
\usepackage{graphicx}
\usepackage{esint}

\makeatletter

\providecommand{\tabularnewline}{\\}

\@ifundefined{textcolor}{}
{%
 \definecolor{BLACK}{gray}{0}
 \definecolor{WHITE}{gray}{1}
 \definecolor{RED}{rgb}{1,0,0}
 \definecolor{GREEN}{rgb}{0,1,0}
 \definecolor{BLUE}{rgb}{0,0,1}
 \definecolor{CYAN}{cmyk}{1,0,0,0}
 \definecolor{MAGENTA}{cmyk}{0,1,0,0}
 \definecolor{YELLOW}{cmyk}{0,0,1,0}
}

\usepackage{babel}

\usepackage{babel}

\makeatother

\usepackage{babel}
\begin{document}

\title{Theory of Interfacial Plasmon-Phonon Scattering in Supported Graphene}

\author{Zhun-Yong Ong}

\email{zhunyong.ong@utdallas.edu}

\selectlanguage{english}%

\affiliation{Department of Materials Science and Engineering, University of Texas
at Dallas RL10, 800 W Campbell Rd RL10, Richardson, TX 75080}

\author{Massimo V. Fischetti}

\email{max.fischetti@utdallas.edu}

\selectlanguage{english}%

\affiliation{Department of Materials Science and Engineering, University of Texas
at Dallas RL10, 800 W Campbell Rd RL10, Richardson, TX 75080}
\begin{abstract}
One of the factors limiting electron mobility in supported graphene
is remote phonon scattering. We formulate the theory of the coupling
between graphene plasmon and substrate surface polar phonon (SPP)
modes, and find that it leads to the formation of interfacial plasmon-phonon
(IPP) modes, from which the phenomena of dynamic anti-screening and
screening of remote phonons emerge. The remote phonon-limited mobilities
for SiO$_{2}$, HfO$_{2}$, h-BN and Al$_{2}$O$_{3}$ substrates
are computed using our theory. We find that h-BN yields the highest
peak mobility, but in the practically useful high-density range the
mobility in HfO$_{2}$-supported graphene is high, despite the fact
that HfO$_{2}$ is a high-$\kappa$ dielectric with low-frequency
modes. Our theory predicts that the strong temperature dependence
of the total mobility effectively vanishes at very high carrier concentrations.
The effects of polycrystallinity on IPP scattering are also discussed. 
\end{abstract}
\maketitle

\section{introduction}

Graphene, a single-layer of hexagonally arranged carbon atoms~\cite{KSNovoselov:Nature05},
has been long considered a promising candidate material for post-Si
CMOS technology and other nano-electronic applications on account
of its excellent electrical~\cite{EHHwang:PRL07} and thermal transport~\cite{AABalandin:NL08}
properties. In suspended single-layer graphene (SLG), the electron
mobility has been demonstrated to be as high as 200,000~cm$^{2}$V$^{-1}$s$^{-1}$~\cite{KIBolotin:SSC08}.
However, in real applications such as a graphene field-effect transistor
(GFET), the graphene is physically supported by an insulating dielectric
substrate such as SiO$_{2}$, and the carrier mobility in such supported-graphene
structures is about one order of magnitude lower~\cite{KIBolotin:SSC08}.
This reduction in carrier mobility is further exacerbated in top-gated
structures in which a thin layer of a high-$\kappa$ dielectric, such
as HfO$_{2}$ or Al$_{2}$O$_{3}$, is deposited or grown on the graphene
sheet~\cite{MCLemme:SSE08,JSMoon:EDL10,JPezoldt:PSS10}. The degradation
of the electrical transport properties is a result of exposure to
environmental perturbations such as scattering by charge traps, surface
roughness, and remote optical phonons which are a kind of surface
excitation. Such environmental effects are encountered in metal-oxide-semiconductor
(MOS) structures~\cite{MVFischetti:JAP01}. Hess and Vogl first suggested
that remote phonons {[}sometimes also known as Fuchs-Kliewer (FK)~\cite{RFuchs:PR65}
surface optical (SO) phonons{]} can have a substantial effect on the
mobility of Si inversion layer carriers~\cite{KHess:SSC79}. Fischetti
and co-workers later studied the effects of remote phonon scattering
in MOS structures and found that high-$\kappa$ oxide layers have
a significant effect on carrier mobility in Si~\cite{MVFischetti:JAP01}
and Ge~\cite{TORegan:JCompElec07}. This method was later applied
by Xiu to study remote phonon scattering in Si nanowires~\cite{KXiu:SISPAD11}.
In Refs.~\cite{MVFischetti:JAP01} and \cite{KXiu:SISPAD11} it was
found that the plasmons in the channel material (Si) hybridized with
the surface polar phonons (SPP) in the nearby dielectric material
to form interfacial plasmon-phonon (IPP) modes. This hybridization
occurrence naturally leads to the screening/anti-screening of the
SPP from the dielectric material. Scattering with these IPP modes
results in a further reduced channel electron mobility in 2D Si and
Si nanowires.

Likewise in supported graphene, remote phonon scattering is one of
the mechanisms believed to reduce the mobility of supported graphene,
with the form of the scattering mechanism varying with the material
properties of the dielectric substrate. Experimentally, hexagonal
boron nitride (h-BN) has been found to be a promising dielectric material
for graphene, and it is commonly believed that this is at least partially
due to the fact that remote phonon scattering is weak with a h-BN
substrate~\cite{CRDean:NatureNanotech10}. On other substrates such
as SiO$_{2}$~\cite{JHChen:NatureNanotech08,VEDorgan:APL10} and
SiC~\cite{JARobinson:NL09,PSutter:NatureMat09}, the mobility of
supported graphene is lower. Thus, it is important to develop an accurate
understanding of remote phonon scattering in order to find an optimal
choice of substrate that will minimize the degradation of carrier
mobility in supported graphene. 

Although the subject of remote phonon scattering in graphene~\cite{SFratini:PRB08,SVRotkin:NL09,VPerebeinos:NL08,AKonar:PRB10,JKViljas:PRB10}
and carbon nanotubes~\cite{VPerebeinos:NL08,SVRotkin:NL09}) has
been broached in the recent past, the basic approach used in the aforementioned
works does not deal adequately with the \emph{dynamic} screening of
the SPP modes. In graphene, dynamic screening of SPP modes has its
origin in SPP-plasmon coupling, and the two time-dependent phenomena
have to be treated within the same framework. Typically, the coupling
phenomenon is ignored, and screening of the SPP modes is approximated
with a Thomas-Fermi (TF) type of static screening~\cite{AKonar:PRB10,XLi:APL10,SFratini:PRB08},
which is adequate for the case of impurity scattering~\cite{SAdam:SSC09,SAdam:PNAS07}
but can lead to a miscalculation of the scattering rates since the
use of static screening underestimates the electron-phonon coupling
strength~\cite{SFratini:PRB08}, especially for higher-frequency
modes. The failure to incorporate correctly SPP-plasmon coupling into
the approach means that the dispersion relation of the SPP (or, more
accurately, of the IPP) modes is incorrect and that the dynamic screening
of the remote phonons is not accounted for in a natural manner. 

To understand the screening phenomenon in our situation, let us first
give a bird's eye view of the physical picture. This picture is somewhat
different from what is found in the more familiar semiconductor-inversion-layer/high-$\kappa$-dielectric
geometry, since the absence of a gap in bulk graphene renders its
dielectric response stronger and qualitatively different -- almost
metal-like, as testified by the presence of Kohn anomalies in the
phonon spectra~\cite{MLazzeri:PRL06,SPisana:NatureMaterials07} --
than the response of a two-dimensional electron gas. Graphene plasmons
interact with the SPP modes through the time-dependent electric field
generated by the latter, and the former are forced to oscillate at
the frequency of the latter ($\omega$). When $\omega$ is less than
the natural frequency of the plasmon ($\omega_{p}$), \emph{i.e.}
$\omega<\omega_{p}$ , the electrons can respond to the SPP mode and
screen its electric field. On the other hand, when $\omega>\omega_{p}$,
the motion of the plasmons lags that of the SPP mode, resulting in
poor or no screening, or even in anti-screening, which can actually
augment the scattering field~\cite{BKRidley:Book99}. In bulk SiO$_{2}$,
the main TO-phonon frequencies are around 56 and 138 meV. At long
wavelengths ($\lambda>10^{-8}$m), the plasmon frequencies for a carrier
density of $10^{12}\mathrm{cm^{-2}}$ are comparable or smaller than
the TO-phonon frequencies. Thus, a TF-type approximation is inadequate
especially for describing the screening (or more accurately, the \emph{anti-screening})
of the 138 meV TO phonon modes. Our calculations suggest that, contrary
to what is found in the semiconductor/high-$\kappa$ case~\cite{MVFischetti:JAP01}
and to the claims made in Ref.~\cite{SFratini:PRB08}, the higher-frequency
SPP modes cannot be ignored despite their reduced Bose-Einstein occupation
factors at room temperature.

It is our intention in this paper to provide a systematic description
of the coupling between the substrate SPP and the graphene plasmon
modes, and relate this coupling to the dynamic screening phenomenon.
Our theory can be generalized to graphene heterostructure such as
double-gated graphene although this falls outside the scope of our
paper and will be the subject of a future work. We begin by deriving
our model of the IPP system. Its dispersion is then calculated from
the model. The pure SO phonon and graphene plasmon branches are compared
with the IPP branches. Also, we compute the electron-IPP and the electron-SPP
coupling coefficients for different substrates (SiO$_{2}$, h-BN,
HfO$_{2}$ and Al$_{2}$O$_{3}$). We show that the IPP modes can
be interpreted as dynamically screened SPP modes. Scattering rates
are then calculated and used to compute the remote phonon-limited
mobility$\mu_{RP}$ for different substrates at room temperature (300
K) with varying carrier density. The temperature dependence of $\mu_{RP}$
at low and high carrier densities is compared. Using the$\mu_{RP}$
results, we analyze the suitability of the various dielectric materials
for use as substrates or gate insulators in nanoelectronics applications.
We also discuss the effects of polycrystallinity on remote phonon
scattering.

\section{Model}

\subsection{Coupling between substrate polar phonons and graphene plasmons}

\begin{figure}
\includegraphics[width=4in]{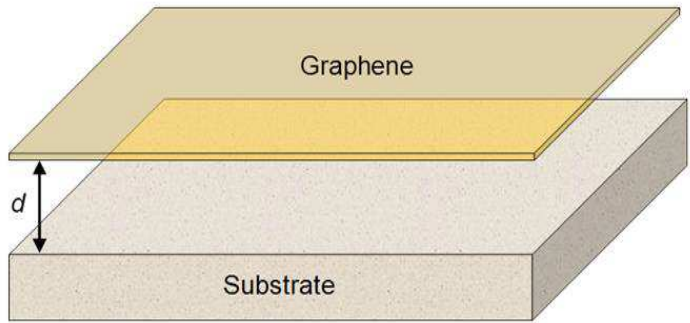}

\caption{Schematic of set up of graphene-substrate system. The SLG is modeled
as a infinitely thin (in the $z$-direction) layer of polarizable
charge. A gap of $d$ separates the graphene charge sheet and the
substrate surface.}

\label{Fig:GrapheneSubstrateSetup} 
\end{figure}

Our approach to constructing the theoretical model of the coupled
plasmon-phonon systems follows closely that of Fischetti, Neumayer
and Cartier~\cite{MVFischetti:JAP01} although some modifications
are needed to describe the plasmon-phonon coupling. One of the primary
difficulties in describing the coupled system is the anisotropy in
the dielectric response of graphene: graphene is polarizable in the
plane but its out-of-plane response is presumably negligible. If the
graphene sheet is modeled as a slab of finite thickness with a dynamic
dielectric response in the in-plane direction {[}$\epsilon_{gr}^{\parallel}(\omega)=\epsilon_{gr}(1-\omega_{p}^{2}/\omega^{2})$
where $\omega_{p}$ is the plasma frequency{]} and none in the out-of-plane
direction {[}$\epsilon_{gr}^{\perp}(\omega)=\mathrm{constant}${]},
the dispersion of the SPPs remains unchanged, indicating that the
SPP and plasmon modes are uncoupled. This absence of coupling is implicitly
assumed in much of the current literature on SPP scattering in graphene~\cite{SFratini:PRB08,XLi:APL10,AKonar:PRB10,JKViljas:PRB10,VPerebeinos:PRB10}
although it has already been shown to be untrue in 2D Si~\cite{MVFischetti:JAP01}
and Si nanowires~\cite{KXiu:SISPAD11}. Furthermore, there is considerable
experimental support for the coupling of graphene plasmons to the
SPPs~\cite{YLiu:PRB08,YLiu:PRB10,ZFei:NL11,RJKoch:PRB10}. As we
will show later, accounting for this coupling results in the formation
of IPP modes which are screened/anti-screened and scatter charge carriers
in graphene more weakly/strongly than the unhybridized SPP modes.
This `uncoupling problem' persists even when one inserts a vacuum
region between the graphene slab and the substrate. Ultimately, this
alleged lack of coupling can be traced back to the continuity of the
electric displacement field $\mathbf{D}$ at the interface between
the graphene slab and the substrate/vacuum. Given that the dynamic
response of the graphene is only in the in-plane directions and that
the coupling should be with the \emph{p}-polarized waves of the substrate,
the slab approach is not likely to be correct. To overcome this difficulty,
we find it is necessary to treat the graphene as a polarizable charge
sheet (as shown in Fig.~\ref{Fig:GrapheneSubstrateSetup}) rather
than as a finite slab with a particular in-plane dielectric function.
This polarization charge then generates a discontinuity in the electric
displacement along the surface of the graphene. It is this discontinuity
that couples the dielectric response of the substrate to that of the
graphene sheet. The basic setup is shown in Fig.~\ref{Fig:GrapheneSubstrateSetup}.
The graphene is an infinitely thin sheet co-planar with the $x$-$y$
plane and floating at a height $d$ above the substrate which occupies
the semi-infinite region $z<0$. Notation-wise, we try to follow Ref.~\cite{MVFischetti:JAP01}.
In the direction perpendicular to the interface, the (ionic) dielectric
response of the substrate is assumed to be due to two optical phonon
modes, an approximation used in Ref.~\cite{MVFischetti:JAP01}, that
is: 
\begin{equation}
\epsilon_{ox}(\omega)=\epsilon_{ox}^{\infty}+(\epsilon_{ox}^{i}-\epsilon_{ox}^{\infty})\frac{\omega_{TO2}^{2}}{\omega_{TO2}^{2}-\omega^{2}}+(\epsilon_{ox}^{0}-\epsilon_{ox}^{i})\frac{\omega_{TO1}^{2}}{\omega_{TO1}^{2}-\omega^{2}}\label{Eq:DielectricEquation}
\end{equation}
 where $\omega_{TO1}$ and $\omega_{TO2}$ are the first and second
transverse optical (TO) angular frequencies (with $\omega_{TO1}<\omega_{TO2}$),
and $\epsilon_{ox}^{\infty}$, $\epsilon_{ox}^{i}$ and $\epsilon_{ox}^{0}$
are the optical, intermediate and static permittivities. We can also
express $\epsilon_{ox}(\omega)$ in the generalized Lyddane-Sachs-Teller
form: 
\[
\epsilon_{ox}(\omega)=\epsilon_{ox}^{\infty}\frac{(\omega_{LO2}^{2}-\omega^{2})(\omega_{LO1}^{2}-\omega^{2})}{(\omega_{TO2}^{2}-\omega^{2})(\omega_{TO1}^{2}-\omega^{2})}
\]
 where $\omega_{LO1}$ and $\omega_{LO2}$ are the first and second
longitudinal optical angular frequencies. The variables $\mathbf{Q}$
and $\mathbf{R}$ represent the two-dimensional wave and coordinate
vector in the $(x,y)$ plane of the interface, respectively.

As in Ref.~\cite{MVFischetti:JAP01}, we try to derive the longitudinal
electric eigenmodes of the system since the transverse modes (given
by poles of the total electric response) correspond to a vanishing
electric field and so to a vanishing coupling with the graphene carriers.
In effect, the longitudinal modes are the transverse-magnetic (TM)
solutions of Maxwell's equations. It was also shown in Ref.~\cite{MVFischetti:JAP01}
that one may ignore the effects of retardation. Therefore, we need
only to employ simpler electrostatics instead of the full Maxwell's
equations.

We begin our derivation by writing down the Poisson equation for the
\emph{bare} scalar potential $\Phi$, 
\begin{equation}
-\nabla^{2}\Phi(\mathbf{R},z)=\frac{1}{\epsilon_{0}}\rho_{ox}(\mathbf{R},z,t),\label{Eq:BarePoisson}
\end{equation}
 where $\rho_{ox}$ is the (periodic) polarization charge distribution
at the surface of the substrate that is the source of scattering,
and $\epsilon_{0}$ is the permittivity of vacuum. Equation~(\ref{Eq:BarePoisson})
describes the electrostatic potential within the graphene. However,
the \emph{effective} scalar potential felt by the graphene carriers
is different and should include the collective screening effect of
the induced electrons/holes, which changes the RHS of Eq.~(\ref{Eq:BarePoisson}).
Hence, we modify Eq.~(\ref{Eq:BarePoisson}) by adding a screening
charge term on its RHS, and we obtain the Poisson equation for the
\emph{screened} scalar potential $\Phi_{scr}$, 
\begin{equation}
-\nabla^{2}\Phi_{scr}(\mathbf{R},z,t)=\frac{1}{\epsilon_{0}}\left[\rho_{ox}(\mathbf{R},z,t)+\rho_{scr}(\mathbf{R},z,t)\right],\label{Eq:EffectivePoisson}
\end{equation}
 where $\rho_{scr}$ is the screening charge term. The integral form
of Eq.~(\ref{Eq:EffectivePoisson}) is: 
\begin{equation}
\Phi_{scr}(\mathbf{R},z,t)=\Phi(\mathbf{R},z,t)+\int d\mathbf{R}'dz'G(\mathbf{R}z,\mathbf{R}'z')\rho_{scr}(\mathbf{R}'z',t)\label{Eq:EffectivePoissonIntegral}
\end{equation}
 where $G(\mathbf{R}z,\mathbf{R}'z')$ is the Green function that
satisfies the boundary conditions {[}see Eqs.~(\ref{Eq:GreenFunctionInterfaceBC}){]},
and the equation: 
\begin{equation}
-\nabla^{2}\left[\epsilon(\mathbf{R},z)G(\mathbf{R}z,\mathbf{R}'z')\right]=\delta(\mathbf{R}-\mathbf{R}',z-z')\ .\label{Eq:RealSpaceGreenFn}
\end{equation}
 The bare potential $\Phi(\mathbf{R},z,t)$ is defined as $\Phi(\mathbf{R},z,t)=\int d\mathbf{R}'dz'G(\mathbf{R}z,\mathbf{R}'z')\rho_{ox}(\mathbf{R}'z',t)$.
The second term on the RHS of Eq.~(\ref{Eq:EffectivePoissonIntegral})
represents the screening charge distribution. The bare and screened
potentials can be written as sums of their Fourier components: \begin{subequations}
\begin{equation}
\Phi(\mathbf{R},z,t)=\sum_{\mathbf{Q}}\phi_{Q,\omega}(z)e^{-i\mathbf{Q}\cdot\mathbf{R}}e^{i\omega t}\ ,
\end{equation}

\begin{equation}
\Phi_{scr}(\mathbf{R},z,t)=\sum_{\mathbf{Q}}\phi_{Q,\omega}^{scr}(z)e^{-i\mathbf{Q}\cdot\mathbf{R}}e^{i\omega t}\ 
\end{equation}
 \label{Eq:FourierComponents} \end{subequations} where it must be
understood that only the real part of Eq.~(\ref{Eq:FourierComponents})
is to be taken here and in the following sections. Given the cylindrical
symmetry of the problem, the Fourier components $\phi_{Q,\omega}$
and $\phi_{Q,\omega}^{scr}$ depend only on the magnitude of the wave
vector $\mathbf{Q}$.

From Eq.~(\ref{Eq:EffectivePoissonIntegral}) we obtain the following
expression for the $z$-dependent part of the Fourier-transformed
screened potential: 
\begin{equation}
\phi_{Q,\omega}^{scr}(z)e^{i\omega t}=\phi_{Q,\omega}(z)e^{i\omega t}+\int dz'G_{Q}(z,z')\rho_{Q,\omega}^{scr}(z',t)\ .\label{Eq:EffectivePoissonIntegralFourier}
\end{equation}
 Equation~(\ref{Eq:EffectivePoissonIntegralFourier}) is solvable
if the polarization charge $\rho_{Q,\omega}^{scr}$ is expressed as
a function of the screened scalar potential. Here, we assume that
$\rho_{Q,\omega}^{scr}$ responds linearly to $\phi_{Q,\omega}^{scr}$,
and write the screening charge term as: 
\begin{equation}
\rho_{Q,\omega}^{scr}(z,t)=e^{2}\Pi(Q,\omega)f(z)\phi_{Q,\omega}^{scr}(z)e^{i\omega t}\label{Eq:ScreeningCharge}
\end{equation}
 where $\Pi(Q,\omega)$ is the in-plane 2D polarization charge term,
and $f(z)$ governs the polarization charge distribution in the out-of-plane
direction. For convenience, we model the graphene as an infinitely
thin sheet of polarized charge and set $f(z)=\delta(z-d)$. Combining
Eqs.~(\ref{Eq:EffectivePoissonIntegralFourier}) and (\ref{Eq:ScreeningCharge}),
we obtain the expression: 
\begin{equation}
\phi_{Q,\omega}^{scr}(z)=\phi_{Q,\omega}(z)+e^{2}\int dz'G_{Q}(z,z')\Pi(Q,\omega)f(z')\phi_{Q,\omega}^{scr}(z')\ .\label{Eq:EffectivePoissonIntegralFourierFinal}
\end{equation}
 The expression in Eq.~(\ref{Eq:EffectivePoissonIntegralFourierFinal})
becomes: 
\begin{align*}
\phi_{Q,\omega}^{scr}(z) & =\phi_{Q,\omega}(z)+e^{2}G_{Q}(z,d)\Pi(Q,\omega)\phi_{Q,\omega}^{scr}(d)\\
 & =\phi_{Q,\omega}(z)+e^{2}G_{Q}(z,d)\Pi(Q,\omega)\phi_{Q,\omega}(d)+e^{4}G_{Q}(z,d)\Pi(Q,\omega)G_{Q}(d,d)\Pi(Q,\omega)\phi_{Q,\omega}^{scr}(d)\\
 & =\ldots\\
 & =\phi_{Q,\omega}(z)+\frac{e^{2}G_{Q}(z,d)\Pi(Q,\omega)}{1-e^{2}G_{Q}(d,d)\Pi(Q,\omega)}\phi_{Q,\omega}(d)
\end{align*}
 and the corresponding component of the electric field perpendicular
to the interface at $z=0$ is: 
\begin{equation}
\hat{\mathbf{z}}\cdot\mathbf{E}_{Q,\omega}\Big|_{z=0}=-\frac{\partial}{\partial z}\phi_{Q,\omega}^{scr}(z)\Bigg|_{z=0}=-\frac{\partial\phi_{Q,\omega}(z,t)}{\partial z}-\frac{\partial G_{Q}(z,d)}{\partial z}\frac{e^{2}\Pi(Q,\omega)}{1-e^{2}G_{Q}(d,d)\Pi(Q,\omega)}\phi_{Q,\omega}(d)\Bigg|_{z=0}\ .
\end{equation}
 For notational simplicity, we write: \begin{subequations} 
\begin{equation}
\phi_{Q,\omega}^{scr}(z)=\phi_{Q,\omega}(z)+G_{Q}(z,d)\mathcal{P}_{Q,\omega}\phi_{Q,\omega}(d)\label{Eq:ScreenedScalar}
\end{equation}
 
\begin{equation}
\hat{\mathbf{z}}\cdot\mathbf{E}_{Q,\omega}=-\frac{\partial\phi_{Q,\omega}(z,t)}{\partial z}-\frac{\partial G_{Q}(z,d)}{\partial z}\mathcal{P}_{Q,\omega}\phi_{Q,\omega}(d)\ ,
\end{equation}
 \end{subequations} where 
\begin{equation}
\mathcal{P}_{Q,\omega}=\frac{e^{2}\Pi(Q,\omega)}{1-e^{2}G_{Q}(d,d)\Pi(Q,\omega)}\ .\label{Eq:Polarization}
\end{equation}
 Here, we emphasize that Eq.~(\ref{Eq:ScreenedScalar}) is the key
to determining the dispersion relation as we shall show later.

The Green function $G_{Q}(z,z')$ in Eq.~(\ref{Eq:EffectivePoissonIntegralFourier})
obeys the relation: 
\begin{equation}
-\left(\frac{\partial^{2}}{\partial z^{2}}-Q^{2}\right)G_{Q}(z,z')=\frac{1}{\epsilon_{0}}\delta(z-z')\ .\label{Eq:GreenFnPoissonEqn}
\end{equation}
 We require the Green function to satisfy the following conditions
at and away from the interface ($z=0$). \begin{subequations} 
\begin{equation}
\epsilon_{0}\frac{dG_{Q}(z=0^{+},z')}{dz}=\epsilon_{ox}^{\infty}\frac{dG_{Q}(z=0^{-},z')}{dz}\label{Eq:GreenFnContinuity-1}
\end{equation}
 
\begin{equation}
G_{Q}(z=0^{+},z')=G_{Q}(z=0^{-},z')\label{Eq:GreenFnContinuity-2}
\end{equation}

\begin{equation}
G_{Q}(z<0,d)=G_{Q}(0,d)e^{+Qz}
\end{equation}

\begin{equation}
G_{Q}(z>d,d)=G_{Q}(d,d)e^{-Q(z-d)}
\end{equation}
 \label{Eq:GreenFunctionInterfaceBC} \end{subequations} The solution
to Eq.~(\ref{Eq:GreenFnPoissonEqn}) is~\cite{JDJackson:Book99}:
\begin{equation}
G_{Q}(z,z')=\left\{ \begin{array}{ll}
\frac{1}{2\epsilon_{0}Q}\left(e^{-Q|z-z'|}-\lambda e^{-Q|z+z'|}\right) & ,z>0\\
\frac{1}{2\epsilon_{0}Q}\left(1-\lambda\right)e^{-Q|z-z'|} & ,z\leq0
\end{array}\right.\label{Eq:FourierGreenFunction}
\end{equation}
 where 
\[
\lambda=\frac{\epsilon_{ox}^{\infty}-\epsilon_{0}}{\epsilon_{ox}^{\infty}+\epsilon_{0}}\ .
\]

The bare potential in Eq.~(\ref{Eq:ScreenedScalar}) can be written
as: 
\[
\phi_{Q,\omega}(z)=\left\{ \begin{array}{ll}
A_{1}e^{-Qz} & ,\ z>0\\
A_{2}e^{+Qz} & ,\ z\leq0
\end{array}\right.
\]
 where $A_{1}$ and $A_{2}$ are the amplitudes of the bare potential
for $z>0$ and $z\leq0$ respectively. Thus, the expression for the
screened potential in Eq.~(\ref{Eq:ScreenedScalar}) is: 
\begin{equation}
\phi_{Q,\omega}^{scr}(z)=\left\{ \begin{array}{ll}
A_{1}e^{-Qz}+G_{Q}(z,d)\mathcal{P}_{Q,\omega}A_{1}e^{-Qd} & ,z>0\\
A_{2}e^{+Qz}+G_{Q}(z,d)\mathcal{P}_{Q,\omega}A_{1}e^{-Qd} & ,z\leq0
\end{array}\right.\ .\label{Eq:ScreenedScalarField}
\end{equation}
 At the interface $z=0$, the continuity of the component of the electric
field parallel to the interface requires the continuity of $\phi_{Q,\omega}^{scr}$,
\emph{i.e.} $\phi_{Q,\omega}^{scr}(z=0^{+})=\phi_{Q,\omega}^{scr}(z=0^{-})$,
giving us: 
\begin{equation}
A_{1}+A_{1}G_{Q}(z=0^{+},d)\mathcal{P}_{Q,\omega}e^{-Qd}=A_{2}+A_{1}G_{Q}(z=0^{-},d)\mathcal{P}_{Q,\omega}e^{-Qd}\ .\label{Eq:ContinuityEField}
\end{equation}
 Similarly, the continuity of the perpendicular component of the electric
displacement, \emph{i.e.}, $\epsilon_{0}\frac{d\phi_{Q,\omega}^{scr}(z=0^{+})}{dz}=\epsilon_{ox}(\omega)\frac{d\phi_{Q,\omega}^{scr}(z=0^{-})}{dz}$
leads to: 
\begin{equation}
\epsilon_{0}\bigg[A_{1}-A_{1}\frac{1}{Q}\frac{dG_{Q}(z=0^{+},d)}{dz}\mathcal{P}_{Q,\omega}e^{-Qd}\bigg]=\epsilon_{ox}(\omega)\bigg[-A_{2}-A_{1}\frac{1}{Q}\frac{dG_{Q}(z=0^{-},d)}{dz}\mathcal{P}_{Q,\omega}e^{-Qd}\bigg]\ .\label{Eq:ContinuityEDisplacement}
\end{equation}
 Substituting Eqs.~(\ref{Eq:GreenFnContinuity-1}) and (\ref{Eq:GreenFnContinuity-2})
into Eqs.~(\ref{Eq:ContinuityEField}) and (\ref{Eq:ContinuityEDisplacement}),
we obtain the following relations: \begin{subequations} 
\begin{equation}
A_{1}=A_{2}
\end{equation}
 
\begin{equation}
\epsilon_{0}\left[1-\frac{1}{Q}\frac{\partial G_{Q}(z=0^{+},d)}{\partial z}\mathcal{P}_{Q,\omega}e^{-Qd}\right]=\epsilon_{ox}(\omega)\left[-1-\frac{1}{Q}\frac{\partial G_{Q}(z=0^{-},d)}{\partial z}\mathcal{P}_{Q,\omega}e^{-Qd}\right]\label{Eq:PreSecularEquation}
\end{equation}
 \end{subequations} Rearranging the terms in Eq.~(\ref{Eq:PreSecularEquation})
we obtain: 
\begin{equation}
\epsilon_{0}+\epsilon_{ox}(\omega)+\left[\epsilon_{ox}(\omega)-\epsilon_{ox}^{\infty}\right]G_{Q}(0,d)\mathcal{P}_{Q,\omega}e^{-Qd}=0\ ,\label{Eq:SecularEquation}
\end{equation}
 which can be rewritten as: 
\[
\left[\epsilon_{0}+\epsilon_{ox}(\omega)\right]\left\{ 1-\left[G_{Q}(d,d)-G_{Q}(0,d)e^{-Qd}\right]e^{2}\Pi(Q,\omega)\right\} -\left(\epsilon_{0}+\epsilon_{ox}^{\infty}\right)G_{Q}(0,d)e^{2}\Pi(Q,\omega)e^{-Qd}=0
\]
 or more explicitly: 
\begin{equation}
\left[\epsilon_{ox}(\omega)+\epsilon_{0}\right]\left[1-\left(1-e^{-2Qd}\right)\frac{e^{2}\Pi(Q,\omega)}{2\epsilon_{0}Q}\right]-\frac{e^{2}\Pi(Q,\omega)}{Q}e^{-2Qd}=0\ .\label{Eq:SecularEquation-1}
\end{equation}
 Equation~(\ref{Eq:SecularEquation}) gives us the dispersion of
the coupled plasmon-phonon modes, and is sometimes called the secular
equation~\cite{MVFischetti:JAP01}. Physically, we expect three branches
(two phonon and one plasmon). We write the coupled plasmon-phonon
modes as $\omega_{Q}^{(1)}$, $\omega_{Q}^{(2)}$ and $\omega_{Q}^{(3)}$
for each $\mathbf{Q}$-point. In the limit $d\rightarrow\infty$,
Eq.~(\ref{Eq:SecularEquation-1}) becomes: 
\[
\left[\epsilon_{ox}(\omega)+\epsilon_{0}\right]\left[1-\frac{e^{2}\Pi(Q,\omega)}{2\epsilon_{0}Q}\right]=0\ 
\]
 which gives us as expected the dispersion for the two uncoupled SPP
branches and the single plasmon branch in isolated graphene.

\subsection{Plasmon and phonon content}

The solutions of Eq.~(\ref{Eq:SecularEquation-1}) represent excitations
of the IPP modes. However, the effective scattering amplitude of a
particular mode may not be substantial if it is plasmon-like. Scattering
with a plasmon-like excitation does not necessarily lead to loss of
momentum since the momentum is simply transferred to the constituent
carriers of the plasmon excitation and there is no change in the total
momentum of all the carriers. On the other hand, scattering with a
phonon-like excitation does lead to a loss of momentum since phonons
belong to a different set of degrees of freedom. Therefore, as in
Ref.~\cite{MVFischetti:JAP01}, it is necessary to define the \emph{phonon
content}~\cite{MEKim:PRB78} of each IPP mode. The phonon content
quantifies the modal fraction that is phonon-like and modulates its
scattering strength. Likewise, we can also define the plasmon content
of the mode. To find the plasmon content, we first consider the two
solutions $\omega_{Q}^{(-g,\alpha)}$ ($\alpha=1,2$) obtained from
the secular equation Eq.~(\ref{Eq:SecularEquation-1}) by ignoring
the polarization response {[}setting $\Pi(Q,\omega)=0${]}. Following
Ref.~\cite{MVFischetti:JAP01}, the plasmon content of the IPP mode
$\omega_{Q}^{(i)}$ is defined here as: 
\begin{equation}
\Phi^{(g)}(\omega_{Q}^{(i)})=\left|\frac{(\omega_{Q}^{(i)2}-\omega_{Q}^{(-g,1)2})(\omega_{Q}^{(i)2}-\omega_{Q}^{(-g,2)2})}{(\omega_{Q}^{(i)2}-\omega_{Q}^{(j)2})(\omega_{Q}^{(i)2}-\omega_{Q}^{(k)2})}\right|\label{Eq:PlasmonContent}
\end{equation}
 where the indices $(i,j,k)$ are cyclical. Note that the expected
`sum rule'~\cite{MVFischetti:JAP01} 
\begin{equation}
\sum_{i=1}^{3}\Phi^{(g)}(\omega_{Q}^{(i)})=1\ ,\label{Eq:PlasmonContentSumRule}
\end{equation}
 holds. Equation~(\ref{Eq:PlasmonContentSumRule}) implies that the
total plasmon weight of the three solutions is equal to one (as it
would be without hybridization). The (non-plasmon) phonon content
is then defined as $1-\Phi^{(g)}(\omega_{Q}^{(i)})$. In order to
distinguish the phonon-1 and phonon-2 parts of the non-plasmon content,
we need to define the relative individual phonon content. For phonon-1,
this is accomplished by ignoring its response and replacing $\epsilon_{ox}(\omega)$
in Eq.~(\ref{Eq:SecularEquation-1}) with $\epsilon_{ox}^{\infty}(\omega_{LO2}^{2}-\omega^{2})/(\omega_{TO2}^{2}-\omega^{2})$.
From the solutions of the modified secular equation ($\omega_{Q}^{(-TO1,1)}$
and $\omega_{Q}^{(-TO1,2)}$), the relative phonon-1 content of mode
$i$ will be: 
\begin{equation}
R^{(TO1)}(\omega_{Q}^{(i)})=\left|\frac{(\omega_{Q}^{(i)2}-\omega_{Q}^{(-TO1,1)2})(\omega_{Q}^{(i)2}-\omega_{Q}^{(-TO1,2)2})}{(\omega_{Q}^{(i)2}-\omega_{Q}^{(j)2})(\omega_{Q}^{(i)2}-\omega_{Q}^{(k)2})}\right|\label{Eq:Phonon1RelativeContent}
\end{equation}
 where, as before, $i$, $j$ and $k$ are cyclical. The relative
phonon-2 content can be similarly defined by replacing the superscript
$(-TO1,\alpha)$ with $(-TO2,\alpha)$. Hence, the TO-phonon-1 content
will be: 
\begin{equation}
\Phi^{(TO1)}(\omega_{Q}^{(i)})=\frac{R^{(TO1)}(\omega_{Q}^{(i)})}{R^{(TO1)}(\omega_{Q}^{(i)})+R^{(TO2)}(\omega_{Q}^{(i)})}\left[1-\Phi^{(g)}(\omega_{Q}^{(i)})\right]\ .\label{Eq:Phonon1Content}
\end{equation}
 The TO-phonon-2 content $\Phi^{(TO2)}(\omega_{Q}^{(i)})$ can be
similarly defined. Given Eqs.~(\ref{Eq:PlasmonContent}) and (\ref{Eq:Phonon1Content}),
the following sum rules have been numerically verified: \begin{subequations}
\[
\sum_{i=1}^{3}\Phi^{(TO1)}(\omega_{Q}^{(i)})=\sum_{i=1}^{3}\Phi^{(TO2)}(\omega_{Q}^{(i)})=1
\]
 
\[
\Phi^{(g)}(\omega_{Q}^{(i)})+\Phi^{(TO1)}(\omega_{Q}^{(i)})+\Phi^{(TO2)}(\omega_{Q}^{(i)})=1
\]
 \label{Eq:PhononContentSumRules} \end{subequations} for each mode
$\omega_{Q}^{(i)}$.

\subsection{Scattering strength}

As we have seen earlier, the IPP modes that result from the SPP-plasmon
coupling have a different dispersion from that of the uncoupled SPP
and plasmon modes. The electric field generated by the IPP modes is
also different from that of the uncoupled SPP and plasmon modes. Since
the remote phonon-electron coupling is derived from the quantization
of the energy density of the electric field~\cite{MVFischetti:JAP01},
we expect this difference in the electric field to be reflected in
the scattering strength of the IPP modes.

To find the scattering strength of an IPP mode, we have to determine
the amplitude of its electric field. In Eq.~(\ref{Eq:ScreenedScalarField})
there are three unknowns ($A_{1}$, $A_{2}$ and $\omega$), two of
which ($A_{1}$ and $\omega$) can only be eliminated through Eqs.~(\ref{Eq:ContinuityEField})
and (\ref{Eq:ContinuityEDisplacement}). To find $A_{1}$, we follow
the semi-classical approach in Ref.~\cite{MVFischetti:JAP01}, where
the time-averaged total energy of the scattering field is set equal
to the zero-point energy. In the following discussion, we set $A_{1}=A_{Q}.$
We first compute the time-averaged electrostatic energy $\langle\mathcal{U}_{Q,\omega}^{scr}\rangle$
associated with the screened field: 
\begin{equation}
\left\langle \mathcal{U}_{Q,\omega}^{scr}\right\rangle =\left\langle \frac{1}{2}\int dzd\mathbf{R}\,\epsilon(\omega_{Q}^{(i)})\left|\nabla\left[\phi_{Q,\omega}^{scr}(z)e^{i\mathbf{Q}\cdot\mathbf{R}-i\omega_{Q}^{(i)}t}\right]\right|^{2}\right\rangle \ .\label{Eq:ScreenedZeroPoint}
\end{equation}
 The angle brackets $\langle\ldots\rangle$ denote time average. The
volume integral in Eq.~(\ref{Eq:ScreenedZeroPoint}) is the result
of three contributions: one from the substrate ($z\leq0$), one from
the graphene-substrate gap ($0<z\leq d$), and one from the region
above the graphene ($z>d$). Each term can be converted into a surface
integral. Adopting a `piecewise approach' to evaluate the integral
in Eq.~(\ref{Eq:ScreenedZeroPoint}), we must evaluate three surface
integrals. To do so, we need the explicit expressions for $G_{Q}(z,d)$:
\begin{equation}
G_{Q}(z,d)=\left\{ \begin{array}{ll}
\frac{1}{2\epsilon_{0}Q}\left(1-\lambda\right)e^{+Q(z-d)} & ,\: z\leq0\\
\frac{1}{2\epsilon_{0}Q}\left[e^{+Q(z-d)}-\lambda e^{-Q(z+d)}\right] & ,\:0<z\leq d\\
\frac{1}{2\epsilon_{0}Q}\left[e^{-Q(z-d)}-\lambda e^{-Q(z+d)}\right] & ,\: z>d
\end{array}\right.
\end{equation}
 and for $-\frac{\partial}{\partial z}G_{Q}(z,d)$: 
\begin{equation}
-\frac{\partial}{\partial z}G_{Q}(z,d)=\left\{ \begin{array}{ll}
-\frac{1}{2\epsilon_{0}}\left(1-\lambda\right)e^{+Q(z-d)} & ,\: z\leq0\\
\frac{1}{2\epsilon_{0}}\left[-e^{+Q(z-d)}-\lambda e^{-Q(z+d)}\right] & ,\:0<z\leq d\\
\frac{1}{2\epsilon_{0}}\left[e^{-Q(z-d)}-\lambda e^{-Q(z+d)}\right] & ,\: z>d
\end{array}\right.
\end{equation}
 We can now evaluate the electrostatic energy in the regions $z\leq0$,
$0<z\leq d$ and $z>d$. 
\begin{equation}
\left\langle \mathcal{U}_{Q,\omega}^{scr}\right\rangle =\left\langle \mathcal{U}_{Q,\omega}^{scr}(z\leq0)\right\rangle +\left\langle \mathcal{U}_{Q,\omega}^{scr}(0<z\leq d)\right\rangle +\left\langle \mathcal{U}_{Q,\omega}^{scr}(z>d)\right\rangle \label{Eq:AllEnergyTerms}
\end{equation}
 As mentioned earlier, the volume integrals in Eq.~(\ref{Eq:ScreenedZeroPoint})
can be recast as surface integrals. Thus, \begin{subequations} 
\begin{equation}
\left\langle \mathcal{U}_{Q,\omega}^{scr}(z\leq0)\right\rangle =\frac{\epsilon_{0}\Omega A_{Q}^{2}Q}{2}\bigg[1-\frac{1}{Q}\frac{\partial G_{Q}(z=0^{+},d)}{\partial z}\mathcal{P}_{Q,\omega}e^{-Qd}\bigg]\bigg[1+G_{Q}(z=0^{+},d)\mathcal{P}_{Q,\omega}e^{-Qd}\bigg]\label{Eq:ZeroPointBottom}
\end{equation}
 
\begin{equation}
\begin{split}\left\langle \mathcal{U}_{Q,\omega}^{scr}(0<z\leq d)\right\rangle =\frac{\overline{\epsilon}_{ox}(\omega_{Q}^{(i)})\Omega A_{Q}^{2}Q}{2}\bigg[1+\frac{1}{Q}\frac{\partial G_{Q}(z=0^{-},d)}{\partial z}\mathcal{P}_{Q,\omega}e^{-Qd}\bigg]\bigg[1+G_{Q}(z=0^{-},d)\mathcal{P}_{Q,\omega}e^{-Qd}\bigg]\\
+\frac{\epsilon_{0}\Omega A_{Q}^{2}Q}{2}\bigg[e^{-Qd}-\frac{1}{Q}\frac{\partial G_{Q}(z=d^{+},d)}{\partial z}\mathcal{P}_{Q,\omega}e^{-Qd}\bigg]\bigg[e^{-Qd}+G_{Q}(z=d^{+},d)\mathcal{P}_{Q,\omega}e^{-Qd}\bigg]
\end{split}
\label{Eq:ZeroPointMiddle}
\end{equation}
 
\begin{equation}
\left\langle \mathcal{U}_{Q,\omega}^{scr}(z>d)\right\rangle =-\frac{\epsilon_{0}\Omega A_{Q}^{2}Q}{2}\bigg[e^{-Qd}-\frac{1}{Q}\frac{\partial G_{Q}(z=d^{-},d)}{\partial z}\mathcal{P}_{Q,\omega}e^{-Qd}\bigg]\bigg[e^{-Qd}+G_{Q}(z=d^{-},d)\mathcal{P}_{Q,\omega}e^{-Qd}\bigg]\ .\label{Eq:ZeroPointTop}
\end{equation}
 \label{Eq:ZeroPointAll} \end{subequations} In Eq.~(\ref{Eq:ZeroPointMiddle}),
$\bar{\epsilon}_{ox}(\omega)$ is the dielectric function of the substrate,
which we distinguish with the overhead bar, and distinct from $\epsilon_{ox}(\omega)$.
As we shall see later, as in Ref.~\cite{MVFischetti:JAP01}, the
function $\bar{\epsilon}_{ox}(\omega)$ is chosen in a way consistent
with the particular excitation that we want. Let us regroup the terms
in Eqs.~(\ref{Eq:ZeroPointAll}) into those on the substrate surface
at $z=0$ and those on the graphene at $z=d$. At $z=0$, we have
\begin{equation}
\begin{split}\left\langle \mathcal{U}_{Q,\omega}^{scr}(z=0)\right\rangle =\frac{\epsilon_{0}\Omega A_{Q}^{2}Q}{2}\bigg[1-\frac{1}{Q}\frac{\partial G_{Q}(z=0^{+},d)}{\partial z}\mathcal{P}_{Q,\omega}e^{-Qd}\bigg]\bigg[1+G_{Q}(z=0^{+},d)\mathcal{P}_{Q,\omega}e^{-Qd}\bigg]\\
+\frac{\overline{\epsilon}_{ox}(\omega_{Q}^{(i)})\Omega A_{Q}^{2}Q}{2}\bigg[1+\frac{1}{Q}\frac{\partial G_{Q}(z=0^{-},d)}{\partial z}\mathcal{P}_{Q,\omega}e^{-Qd}\bigg]\bigg[1+G_{Q}(z=0^{-},d)\mathcal{P}_{Q,\omega}e^{-Qd}\bigg]
\end{split}
\label{Eq:SubstrateSurfaceZeroPoint}
\end{equation}
 while at $z=d$, we have: 
\begin{equation}
\begin{split}\left\langle \mathcal{U}_{Q,\omega}^{scr}(z=d)\right\rangle =\frac{\epsilon_{0}\Omega A_{Q}^{2}Q}{2}\bigg[e^{-Qd}-\frac{1}{Q}\frac{\partial G_{Q}(z=d^{+},d)}{\partial z}\mathcal{P}_{Q,\omega}e^{-Qd}\bigg]\bigg[e^{-Qd}+G_{Q}(z=d^{+},d)\mathcal{P}_{Q,\omega}e^{-Qd}\bigg]\\
-\frac{\epsilon_{0}\Omega A_{Q}^{2}Q}{2}\bigg[e^{-Qd}+\frac{1}{Q}\frac{\partial G_{Q}(z=d^{-},d)}{\partial z}\mathcal{P}_{Q,\omega}e^{-Qd}\bigg]\bigg[e^{-Qd}+G_{Q}(z=d^{-},d)\mathcal{P}_{Q,\omega}e^{-Qd}\bigg]
\end{split}
\label{Eq:GrapheneSurfaceZeroPoint}
\end{equation}
 We have to be careful in computing $\left\langle \mathcal{U}_{Q,\omega}^{scr}\right\rangle $.
Mathematically, it may seem that we ought to set $\left\langle \mathcal{U}_{Q,\omega}^{scr}\right\rangle =\left\langle \mathcal{U}_{Q,\omega}^{scr}(z=0)\right\rangle +\left\langle \mathcal{U}_{Q,\omega}^{scr}(z=d)\right\rangle $.
However, note that the term $\left\langle \mathcal{U}_{Q,\omega}^{scr}\right\rangle $
accounts for the various excitation effects, ionic and electronic,
but the term $\left\langle \mathcal{U}_{Q,\omega}^{scr}(z=d)\right\rangle $
corresponds to the charge singularity in the zero-thickness graphene
sheet. It is a `self-interaction' of the charge distribution in the
graphene which has no dependence on $\omega$, unlike $\left\langle \mathcal{U}_{Q,\omega}^{scr}(z=0)\right\rangle $,
so that we should not expect it to contribute physically to the scattering
of the graphene carriers. To see this more clearly, we rewrite Eq.~(\ref{Eq:ScreenedZeroPoint})
as: 
\[
\left\langle \mathcal{U}_{Q,\omega}^{scr}\right\rangle =\left\langle \frac{1}{2}\int dzd\mathbf{R}\,\mathbf{D}\cdot\mathbf{E}\right\rangle =\left\langle \frac{1}{2}\int dzd\mathbf{R}\,(\epsilon_{0}\mathbf{E}+\mathbf{P}_{L}+\mathbf{P}_{e})\cdot\mathbf{E}\right\rangle 
\]
 where $\mathbf{P}_{L}$ and $\mathbf{P}_{e}$ are the polarization
fields of the lattice (substrate) and the graphene electronic excitation
respectively. The following identification can be made: 
\begin{align*}
\left\langle \mathcal{U}_{Q,\omega}^{scr}(z=0)\right\rangle  & =\left\langle \frac{1}{2}\int dzd\mathbf{R}\,(\epsilon_{0}\mathbf{E}+\mathbf{P}_{L})\cdot\mathbf{E}\right\rangle \\
\left\langle \mathcal{U}_{Q,\omega}^{scr}(z=d)\right\rangle  & =\left\langle \frac{1}{2}\int dzd\mathbf{R}\,\mathbf{P}_{e}\cdot\mathbf{E}\right\rangle 
\end{align*}
 and since we are only interested in the interaction of the \emph{lattice}
polarization field with the graphene carriers, we set $\left\langle \mathcal{U}_{Q,\omega}^{scr}\right\rangle =\left\langle \mathcal{U}_{Q,\omega}^{scr}(z=0)\right\rangle $,
\emph{i.e.}: 
\begin{eqnarray}
\left\langle \mathcal{U}_{Q,\omega}^{scr}\right\rangle  & = & \frac{\Omega A_{Q}^{2}Q}{2}\left[\epsilon_{0}+\bar{\epsilon}_{ox}(\omega_{Q}^{(i)})+\left(\bar{\epsilon}_{ox}(\omega_{Q}^{(i)})-\epsilon_{ox}^{\infty}\right)G_{Q}(z=0,d)\mathcal{P}_{Q,\omega}e^{-Qd}\right]\:\nonumber \\
 &  & \times\left[1+G_{Q}(z=0,d)\mathcal{P}_{Q,\omega}e^{-Qd}\right]\label{Eq:ScreenedElectostaticEnergy}
\end{eqnarray}
 where we have used the relationship $\frac{1}{Q}\frac{\partial}{\partial z}G_{Q}(z=0^{-},d)=G_{Q}(z=0,d)$.
In Eq.~(\ref{Eq:ScreenedElectostaticEnergy}), the first factor $[\epsilon_{0}+\bar{\epsilon}_{ox}(\omega)-\ldots]$
resembles the secular equation, Eq.~(\ref{Eq:SecularEquation}).
Indeed, if we replace $\overline{\epsilon}_{ox}(\omega)$ with $\epsilon_{ox}(\omega)$,
then $\left\langle \mathcal{U}_{Q,\omega}^{scr}\right\rangle =0$
as expected. This is no coincidence since $\mathcal{U}_{Q,\omega}^{scr}$
represents the energy of the charge distribution present at the substrate-vacuum
interface and the secular equation in Eq.~(\ref{Eq:SecularEquation})
is a statement about the \emph{absence} of charges at the substrate-vacuum
interface. This also confirms our earlier choice of excluding the
contribution from the surface charges at $z=d$, since that contribution
does not disappear when we replace $\overline{\epsilon}_{ox}(\omega)$
with $\epsilon_{ox}(\omega)$. By regrouping the terms according to
the position of their charge distribution in Eq.~(\ref{Eq:AllEnergyTerms}),
we make the relationship to the secular equation Eq.~(\ref{Eq:SecularEquation})
manifest. Using Eq.~(\ref{Eq:Polarization}), the expression for
the screened electrostatic energy can be rewritten as: 
\begin{equation}
\left\langle \mathcal{U}_{Q,\omega}^{scr}\right\rangle =\frac{\Omega A_{Q}^{2}Q}{2}\left[\bar{\epsilon}_{ox}(\omega_{Q}^{(i)})-\epsilon_{ox}(\omega_{Q}^{(i)})\right]\left(\frac{1-G_{Q}(d,d)e^{2}\Pi(Q,\omega_{Q}^{(i)})+G_{Q}(0,d)e^{2}\Pi(Q,\omega_{Q}^{(i)})e^{-Qd}}{1-G_{Q}(d,d)e^{2}\Pi(Q,\omega_{Q}^{(i)})}\right)^{2}\ .\label{Eq:ScreenedElectrostaticEnergyNeat}
\end{equation}

We use the relationship $\langle\mathcal{W}_{Q,\omega}^{scr}\rangle=2\langle\mathcal{U}_{Q,\omega}^{scr}\rangle$
to obtain the time-averaged total energy, and set it equal to the
zero-point energy, {\em i.e.}, $\frac{1}{2}\hbar\omega_{Q}^{(i)}=\langle\mathcal{W}_{Q,\omega}^{scr}\rangle$,
so that: 
\[
\frac{1}{2}\hbar\omega_{Q}^{(i)}=\Omega A_{Q}^{2}Q\left[\bar{\epsilon}_{ox}(\omega_{Q}^{(i)})-\epsilon_{ox}(\omega_{Q}^{(i)})\right]\left(\frac{1-G_{Q}(d,d)e^{2}\Pi(Q,\omega_{Q}^{(i)})+G_{Q}(0,d)e^{2}\Pi(Q,\omega_{Q}^{(i)})e^{-Qd}}{1-G_{Q}(d,d)e^{2}\Pi(Q,\omega_{Q}^{(i)})}\right)^{2}\ .
\]
 Therefore, the squared amplitude of the field is: 
\[
A_{Q}^{2}=\frac{\hbar\omega_{Q}^{(i)}}{2\Omega Q\left[\bar{\epsilon}_{ox}(\omega_{Q}^{(i)})-\epsilon_{ox}(\omega_{Q}^{(i)})\right]}\left(\frac{1-G_{Q}(d,d)e^{2}\Pi(Q,\omega_{Q}^{(i)})+G_{Q}(0,d)e^{2}\Pi(Q,\omega_{Q}^{(i)})e^{-Qd}}{1-G_{Q}(d,d)e^{2}\Pi(Q,\omega_{Q}^{(i)})}\right)^{-2}\ .
\]
 To determine the strength of the scattering field, say for the TO1
phonon, we take the difference between (i) the squared amplitude of
the field with the TO1 mode frozen and (ii) that of the field with
the mode in full response. In (i), we set: 
\[
\bar{\epsilon}_{ox}^{TO1,\infty}(\omega)=\epsilon_{ox}^{\infty}\left(\frac{\omega_{LO2}^{2}-\omega^{2}}{\omega_{TO2}^{2}-\omega^{2}}\right)
\]
 and 
\[
\bar{\epsilon}_{ox}^{TO1,0}(\omega)=\epsilon_{ox}^{\infty}\left(\frac{\omega_{LO2}^{2}-\omega^{2}}{\omega_{TO2}^{2}-\omega^{2}}\right)\frac{\omega_{LO1}^{2}}{\omega_{TO1}^{2}}\ .
\]
 The squared amplitude of the TO1 scattering field for $\omega=\omega_{Q}^{(i)}$
is: 
\begin{eqnarray}
A_{TO1}(Q,\omega_{Q}^{(i)})^{2} & = & \frac{\hbar\omega_{Q}^{(i)}}{2\Omega Q}\left(\frac{1}{\bar{\epsilon}_{ox}^{TO1,\infty}(\omega_{Q}^{(i)})-\epsilon_{ox}(\omega_{Q}^{(i)})}-\frac{1}{\bar{\epsilon}_{ox}^{TO1,0}(\omega_{Q}^{(i)})-\epsilon_{ox}(\omega_{Q}^{(i)})}\right)\Phi^{(TO1)}(\omega_{Q}^{(i)})\nonumber \\
 &  & \times\left(\frac{1-G_{Q}(d,d)e^{2}\Pi(Q,\omega_{Q}^{(i)})+G_{Q}(0,d)e^{2}\Pi(Q,\omega_{Q}^{(i)})e^{-Qd}}{1-G_{Q}(d,d)e^{2}\Pi(Q,\omega_{Q}^{(i)})}\right)^{-2}\ .\label{Eq:SqAmpScattField}
\end{eqnarray}
 The expression for the TO2 scattering field can be similarly obtained.
Therefore, the TO1 \emph{effective} scattering field can be written
as: 
\begin{equation}
\phi_{Q,\omega}^{scr}(d)=A_{TO1}(Q,\omega_{Q}^{(i)})\left[e^{-Qd}+G_{Q}(z,d)\frac{e^{2}\Pi(Q,\omega_{Q}^{(i)})}{1-e^{2}G_{Q}(d,d)\Pi(Q,\omega_{Q}^{(i)})}e^{-Qd}\right]\ .\label{Eq:TO1EffScattField}
\end{equation}
 The scattering potential is: 
\begin{equation}
V(\mathbf{R},z)=\sum_{l=1}^{3}\sum_{\mu=1}^{2}eA_{TO1}(Q,\omega_{Q}^{(l)})\left[e^{-Qz}+G_{Q}(z,d)\mathcal{P}_{Q,\omega_{Q}^{(l)}}e^{-Qd}\right]e^{i\mathbf{Q}\cdot\mathbf{R}-i\omega_{Q}^{(l)}t}\left(a_{\mathbf{Q}}^{(l)}+a_{\mathbf{-Q}}^{(l)\dagger}\right)\label{Eq:ScatteringPotential}
\end{equation}
 where $a_{\mathbf{Q}}^{(l)}$ ($a_{\mathbf{Q}}^{(l)\dagger}$) is
the annihilation (creation) operator for the mode corresponding to
$\mathbf{Q}$ and $\omega_{Q}^{(l)}$. Generally, the graphene field
operator can be written in the spinorial form as: 
\begin{equation}
\Psi(\mathbf{R},z)=\frac{1}{\sqrt{2\Omega}}\sum_{s=\pm1}\sum_{\mathbf{K}}\left[\left(\begin{array}{c}
1\\
se^{i\theta_{\mathbf{K}}}
\end{array}\right)c_{\mathbf{K}}^{s\mathcal{K}}+\left(\begin{array}{c}
e^{i\theta_{\mathbf{K}}}\\
s
\end{array}\right)c_{\mathbf{K}}^{s\mathcal{K}'}\right]e^{i\mathbf{K}\cdot\mathbf{R}}\sqrt{\delta(z-d)}\label{Eq:GrapheneFieldOperator}
\end{equation}
 where $\mathcal{K}(\mathcal{K}')$ denotes the $\mathcal{K}(\mathcal{K}')$
valley, and the $+(-)$ sign corresponds to the $\pi(\pi^{*})$ band;
$c_{\mathbf{K}}^{s\mathcal{K}}$($c_{\mathbf{K}}^{s\mathcal{K}\dagger}$)
is the annihilation (creation) operator of the $s$-band $\mathbf{K}$
electron state at the $\mathcal{K}$ valley. Therefore, the interaction
term is 
\[
H_{int}=\int dz\int d\mathbf{R}\Psi^{\dagger}(\mathbf{R},z)V(\mathbf{R},z)\Psi(\mathbf{R},z)
\]
 and, if we neglect the inter-valley terms, simplifies to: 
\begin{equation}
H_{int}\approx\sum_{l=1}^{3}\sum_{s_{1},s_{2}}\sum_{\mathbf{K},\mathbf{Q}}M_{Q}^{(l)}\alpha_{s_{1}\mathbf{K+Q},s_{2}\mathbf{K}}\left(c_{\mathbf{K}+\mathbf{Q}}^{s_{1}\mathcal{K}\dagger}c_{\mathbf{K}}^{s_{2}\mathcal{K}}+s_{1}s_{2}c_{\mathbf{K}+\mathbf{Q}}^{s_{1}\mathcal{K}'\dagger}c_{\mathbf{K}}^{s_{2}\mathcal{K}'}\right)\left(a_{\mathbf{Q}}^{(l)}+a_{\mathbf{-Q}}^{(l)\dagger}\right)\label{Eq:InteractionTerm}
\end{equation}
 where 
\[
\alpha_{s_{1}\mathbf{K_{1}},s_{2}\mathbf{K_{2}}}=\frac{1+s_{1}s_{2}e^{-i(\theta_{\mathbf{K}_{1}}-\theta_{\mathbf{K_{2}}})}}{2}
\]
 is the overlap integral that comes from the inner product of the
spinors, and 
\begin{eqnarray}
M_{Q}^{(l)} & = & \sum_{\mu=1}^{2}\left[\frac{e^{2}\hbar\omega_{Q}^{(l)}}{2\Omega Q}\left(\frac{1}{\bar{\epsilon}_{ox}^{TO\mu,\infty}(\omega_{Q}^{(l)})-\epsilon_{ox}(\omega_{Q}^{(l)})}-\frac{1}{\bar{\epsilon}_{ox}^{TO\mu,0}(\omega_{Q}^{(l)})-\epsilon_{ox}(\omega_{Q}^{(l)})}\right)\Phi^{(TO\mu)}(\omega_{Q}^{(l)})\right]^{1/2}\nonumber \\
 &  & \times\left|1-G_{Q}(d,d)e^{2}\Pi(Q,\omega_{Q}^{(l)})+G_{Q}(0,d)e^{2}\Pi(Q,\omega_{Q}^{(l)})e^{-Qd}\right|^{-1}\label{Eq:ElectronPhononMQ}
\end{eqnarray}
 is the electron-phonon coupling coefficient corresponding to the
$\omega_{Q}^{(l)}$ mode.

\subsection{Landau damping}

At sufficiently short wavelengths, plasmons cease to be proper quasi-particle
excitations because of Landau damping~\cite{BKRidley:Book99}. To
model this phenomenon, albeit approximately, we take that to be the
case when the pure graphene plasmon excitation, whose dispersion $\omega=\omega_{p}(Q)$
is determined by the expression $1-e^{2}G_{Q}(d,d)\Pi(Q,\omega)=0$,
enters the intra-band single-particle excitation (SPE) continuum~\cite{EHHwang:PRB07}.
This happens when the plasmon branch crosses the electron dispersion
curve, \emph{i.e.} when $\omega_{p}=\hbar v_{F}Q$, and the wave vector
at which this happens is $Q_{c}$. When $Q<Q_{c}$, the electron-phonon
coupling coefficient in Eq.~(\ref{Eq:InteractionTerm}) is that of
Eq.~(\ref{Eq:ElectronPhononMQ}). Although the lower-frequency IPP
branches may undergo Landau damping from intra-band SPE as $\omega_{Q}^{(l)}<v_{F}Q$,
we still retain them because the sum rules in Eqs.~(\ref{Eq:PlasmonContentSumRule})
and (\ref{Eq:PhononContentSumRules}) require us to maintain charge
conservation~\cite{BKRidley:Book99}. On the other hand, when $Q>Q_{c}$,
Landau damping is assumed to dominate all the IPP modes and the coupling
between the substrate SPP modes and the graphene plasmons can be ignored.
Instead of scattering with three IPP modes for each given wave vector,
we revert to using only two SPP modes. This allows us to satisfy the
sum rules in Eq.~(\ref{Eq:PhononContentSumRules}). In this case,
the electron-phonon coupling coefficient in Eq.~(\ref{Eq:ElectronPhononMQ})
can be rewritten as: 
\[
M_{Q}^{(l)}=\sum_{\mu=1}^{2}\left[\frac{e^{2}\hbar\omega_{Q}^{(l)}}{2\Omega Q}\left(\frac{1}{\bar{\epsilon}_{ox}^{TO\mu,\infty}(\omega_{Q}^{(l)})+\epsilon_{0}}-\frac{1}{\bar{\epsilon}_{ox}^{TO\mu,0}(\omega_{Q}^{(l)})+\epsilon_{0}}\right)\Phi^{(TO\mu)}(\omega_{Q}^{(l)})\right]^{1/2}
\]
 where $l=\mathrm{SO_{1}},\mathrm{SO_{2}}$ indexes the SPP branch.

\section{Results and discussion}

\subsection{Numerical evaluation}

Having set up the theoretical framework for electron-IPP interaction,
we compute the dispersion of the coupled interfacial plasmon-phonon
modes and study the electrical transport properties.

\subsubsection{Interfacial plasmon-phonon dispersion }

In this section we compute the scattering rates from the remote phonons
by employing the dispersion relation ($\omega_{Q}^{(l)}$) and the
electron-phonon coupling coefficient ($M_{Q}^{(l)}$), which can be
determined by solving Eqs.~(\ref{Eq:SecularEquation}) and (\ref{Eq:InteractionTerm}),
respectively. For simplicity, to solve the latter equations we use
the zero-temperature, long-wavelength approximation for $\Pi(Q,\omega)$~\cite{BWunsch:NJP06,EHHwang:PRB07}:
\begin{equation}
\Pi(Q,\omega)=\frac{Q^{2}E_{F}}{\pi\hbar^{2}\omega^{2}}\label{Eq:LongWavelengthApprox}
\end{equation}
 where $E_{F}$ is the Fermi level which can be determined from the
carrier density $n$ via the relation $n=E_{F}^{2}/(\pi\hbar^{2}v_{F}^{2})$.

\begin{figure}
\includegraphics[width=4in]{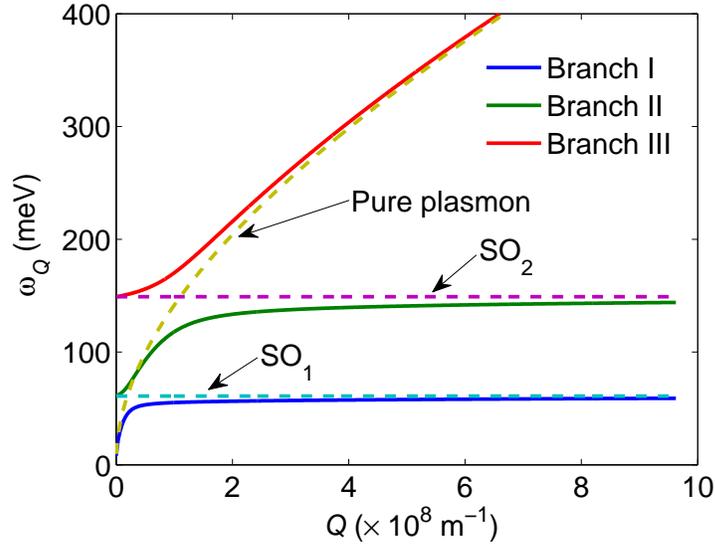}

\caption{Dispersion relation of coupled interfacial plasmon-phonon system with
$n=10^{12}\mathrm{cm^{-2}}$. The three hybrid IPP branches are labeled
I (blue), II (green) and III (red). The uncoupled pure SO phonon (1
and 2) and plasmon branches are drawn with dashed lines. In the limits
$Q\rightarrow0$ and $Q\rightarrow\infty$, the IPP branches converge
to the pure phonon and plasmon branches. In between, they are a mix
of the pure branches.}

\label{Fig:Dispersion1E12} 
\end{figure}

In Fig.~\ref{Fig:Dispersion1E12} we show the dispersion relation
for an SiO$_{2}$ substrate with $n=10^{12}\mathrm{cm^{-2}}$. The
three coupled IPP branches are drawn with solid lines and labeled
I, II and III while the dispersion of the uncoupled modes is drawn
in dashed lines in the figure. The branches labeled `$\mathrm{SO_{1}}$'
(61 meV) and `$\mathrm{SO_{2}}$' (149 meV) have a flat dispersion
and are determined from the quantity: 
\[
\epsilon_{0}+\epsilon_{ox}(\omega)=0
\]
 while the branch labeled `Pure plasmon' is determined from the zeros
of the equation: 
\begin{equation}
1-G_{Q}(d,d)e^{2}\Pi(Q,\omega)\label{Eq:PurePlasmonDispersion}
\end{equation}
 which gives the dispersion of the pure graphene plasmons when the
frequency dependence of the substrate dielectric function is neglected
and only the effect of the substrate image charges is taken into account.
We observe that in the long wavelength limit ($Q\rightarrow0$), branches
I, II and III converge asymptotically to the `pure' plasmon, $\mathrm{SO_{1}}$,
and $\mathrm{SO_{2}}$ branches respectively. On the other end, as
$Q\rightarrow\infty$, branches I, II and III converge asymptotically
to the pure $\mathrm{SO_{1}}$, $\mathrm{SO_{2}}$ and plasmon branches
respectively. At intermediate values of $Q$ the IPP branches are
a mixture of the pure branches. The coupling between pure SO phonons
and graphene plasmons has often been ignored in transport studies
based on the dispersionless unscreened, decoupled SO modes \cite{AKonar:PRB10,JKViljas:PRB10,KZou:PRL10,SFratini:PRB08,SVRotkin:NL09,VPerebeinos:PRB10,XLi:APL10}
On the other hand, using many-body techniques, Hwang, Sensarma and
Das Sarma~\cite{EHHwang:PRB10} have studied the remote phonon-plasmon
coupling in supported graphene and were able to reproduce the coupled
plasmon-phonon dispersion observed by Liu and Willis~\cite{YLiu:PRB08,YLiu:PRB10}
in their angle-resolved electron-energy-loss spectroscopy experiments
on epitaxial graphene grown on SiC. Similar results of strongly coupled
plasmon-phonon modes were reported by Koch, Seyller and Schaefer~\cite{RJKoch:PRB10}.
Fei and co-workers also found evidence of this plasmon-phonon coupling
in the graphene-SiO$_{2}$ system in their infrared nanoscopy experiments~\cite{ZFei:NL11}.
Given the increasing experimental support for the hybridization of
the SPPs with the graphene plasmons, it is interesting to investigate
the effect of these coupled modes on carrier transport in graphene.

\subsubsection{Electron-phonon coupling}

\begin{figure}
\includegraphics[width=4in]{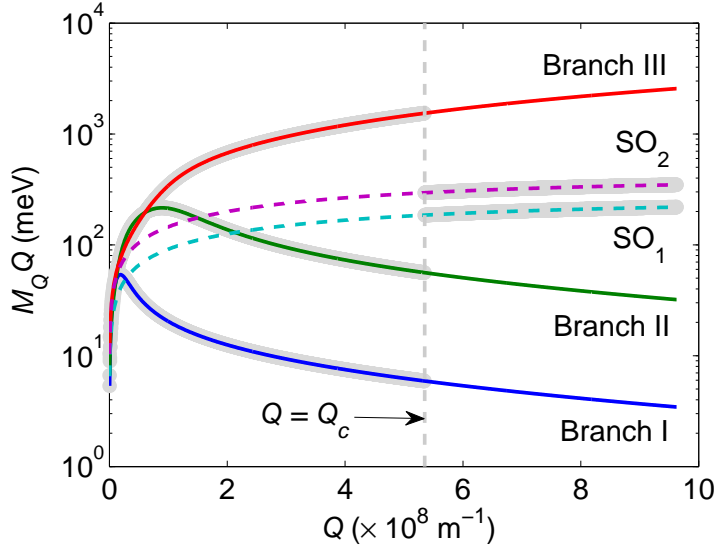}

\caption{Plot of $M_{Q}^{(l)}Q$ for $n=10^{12}\mathrm{cm^{-2}}$ in SiO$_{2}$.
The IPP branches are labeled I (blue), II (green) and III (red), and
the SPP branches are labeled $\mathrm{SO_{1}}$ and $\mathrm{SO_{2}}$.
The cutoff wave vector $Q_{c}$ is drawn in gray dashed lines. When
$Q<Q_{c}$, we use the part of the IPP branches shaded in gray, and
when $Q\geq Q_{c}$, we use the part of the SPP branches shaded in
gray.}

\label{Fig:CouplingCoeffSiO2} 
\end{figure}

Here, the electron-phonon coupling coefficients $M_{Q}^{(l)}$ of
the IPP and the SPP modes are compared. Recall that IPP modes are
formed through the hybridization of the SPP and graphene plasmon modes,
and their coupling to the graphene electrons are different to that
of the SPP modes. It is sometimes assumed~\cite{AKonar:PRB10,SFratini:PRB08}
that the SPP modes are screened by the plasmons, and the IPP-electron
coupling is weaker than the SPP-electron coupling. As we have discussed
above, this assumption does not hold when the frequency of the IPP
mode is higher than the plasmon frequency. We plot the $M_{Q}^{(l)}Q$
for the SPP and IPP modes in Fig.~\ref{Fig:CouplingCoeffSiO2}.We
first notice that at small $Q$, the coupling terms for branches I
and II are actually larger than those for $\mathrm{SO_{1}}$ and $\mathrm{SO_{2}}$,
even though I and II are phonon-like. This is because at long wavelengths,
$\omega_{p}<\omega_{Q}^{(l)}$ for $l=1,2$, resulting in anti-screening,
effect which enhances the SPP electric field. For the plasmon-like
branch III, $M_{Q}^{(III)}$ is actually much larger than the those
for $\mathrm{SO_{1}}$ and $\mathrm{SO_{2}}$ over the entire range
of $Q$ values. When we take Landau damping into account, we use the
coupling coefficients (shaded in gray in Fig.~\ref{Fig:CouplingCoeffSiO2})
of branches I, II and II for $Q<Q_{c}$ and of $\mathrm{SO_{1}}$
and $\mathrm{SO_{2}}$ for $Q\geq Q_{c}$.

\subsection{Substrate-limited mobility}

\begin{figure}
\includegraphics[width=3in]{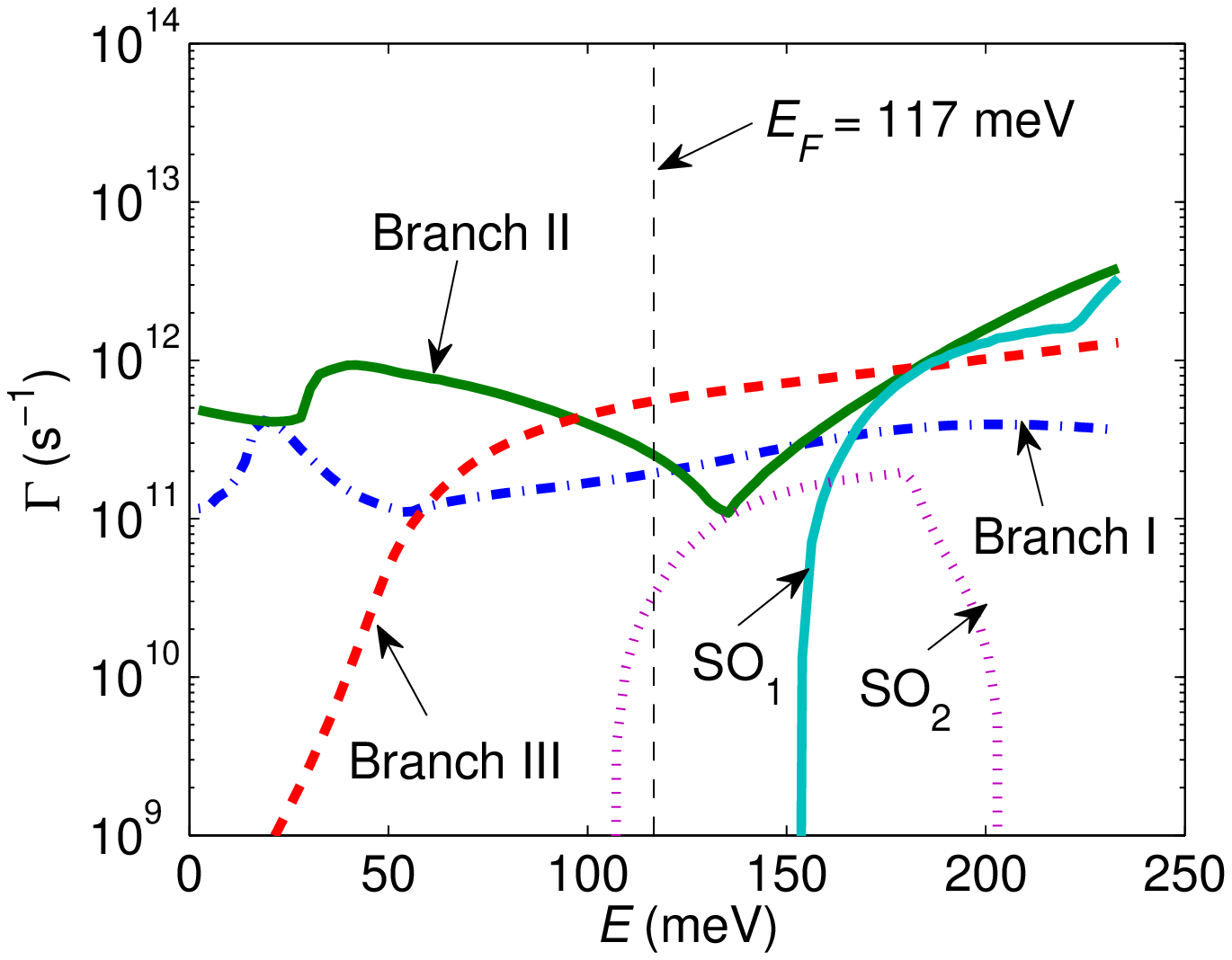} \includegraphics[width=3in]{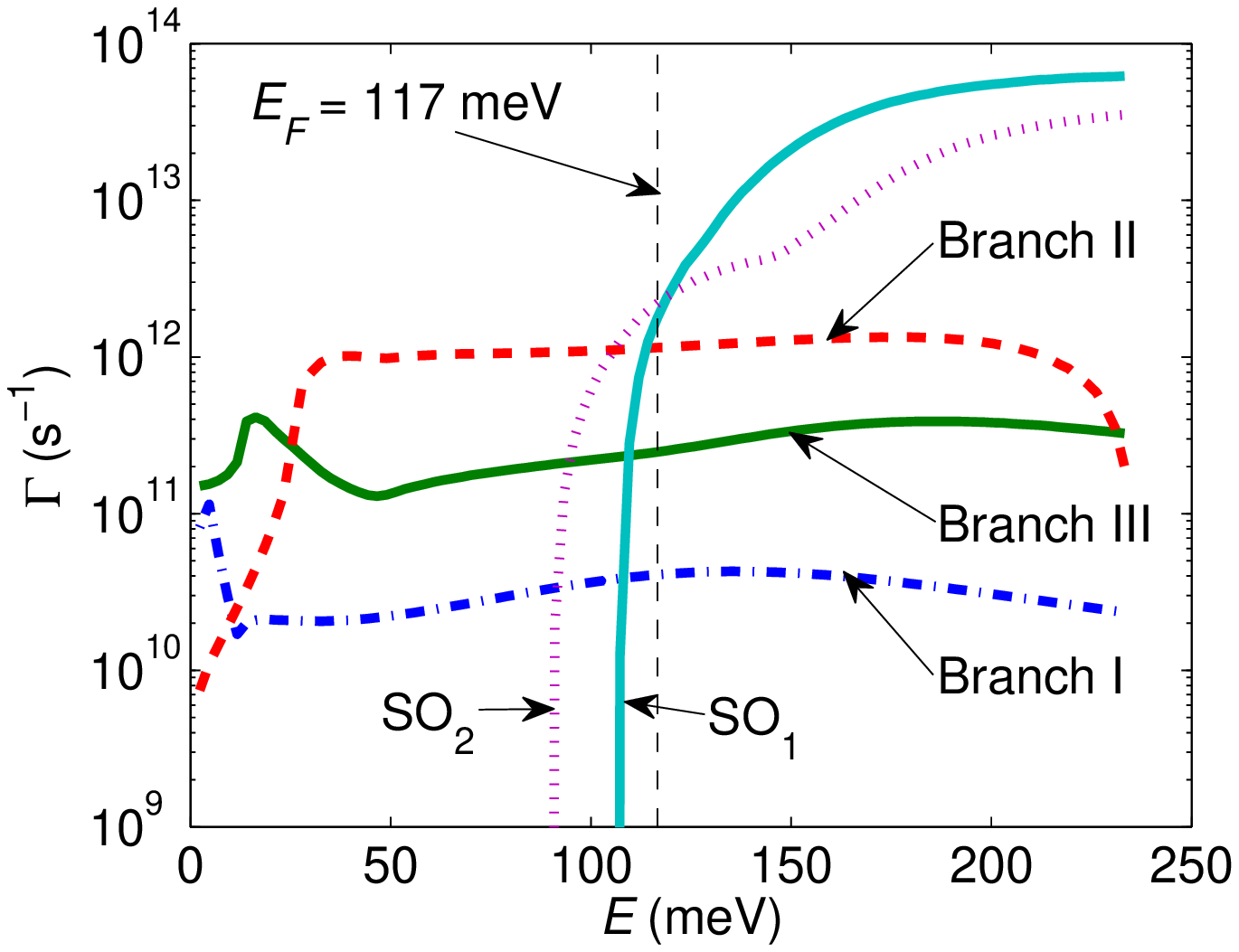}

\caption{Plot of scattering rates at $n=10^{12}\mathrm{\ cm^{-2}}$ for different
substrates: (left) $\mathrm{SiO_{2}}$ and (right) HfO$_{2}$. In
SiO$_{2}$, the plasmon-like branch III dominates the scattering rate
at $E=E_{F}$. In HfO$_{2}$, branches III, $\mathrm{SO_{1}}$ and
$\mathrm{SO_{2}}$ dominate the scattering rate at $E=E_{F}=117\mathrm{\ meV}$.
The SPP branches ($\mathrm{SO_{1}}$ and $\mathrm{SO_{2}}$) do not
contribute much to the Fermi-level scattering rate in SiO$_{2}$ because
of their higher frequencies and smaller occupation factors.}

\label{Fig:ScatteringRateSiO2} 
\end{figure}

The momentum relaxation rate for an electron in band $s$ with wave
vector $\mathbf{K}$ can be written as: 
\begin{align}
\Gamma_{RP}(s,\mathbf{K}) & =\frac{2\pi}{\hbar}\sum_{l}\sum_{s'}\sum_{\mathbf{Q}}\left|M_{Q}^{(l)}\alpha_{s\mathbf{K+Q},s'\mathbf{K}}\right|^{2}\left[1-ss'\cos(\theta_{\mathbf{K+Q}}-\theta_{\mathbf{K}})\right]\nonumber \\
 & \ \times\Bigg\{\left[1+N_{B}(\omega_{Q}^{(l)})\right]\left[1-f(E_{s'\mathbf{K+Q}})\right]\delta(E_{s\mathbf{K}}-E_{s'\mathbf{K+Q}}-\hbar\omega_{Q}^{(l)})\nonumber \\
 & \ +N_{B}(\omega_{Q}^{(l)})\left[1-f(E_{s'\mathbf{K+Q}})\right]\delta(E_{s\mathbf{K}}-E_{s'\mathbf{K+Q}}+\hbar\omega_{Q}^{(l)})\Bigg\}\label{Eq:MomentumRelaxationRate}
\end{align}
 where $N_{B}(\omega)=(e^{\hbar\omega/k_{B}T}-1)^{-1}$, $f(E)=[e^{(E-E_{F})/k_{B}T}+1]^{-1}$
and $E_{s\mathbf{K}}=s\hbar|\mathbf{K|}$. In assuming the latter
expression, we use the Dirac-conical approximation. Equation~(\ref{Eq:MomentumRelaxationRate})
automatically includes the Fermi-Dirac distribution of the final states
and remains applicable when the doping level is high. The individual
scattering rates for the screened (I, II and III) and unscreened ($\mathrm{SO_{1}}$
and $\mathrm{SO_{2}}$) branches at the carrier concentration of $n=10^{12}\mathrm{cm^{-2}}$
in SiO$_{2}$ and HfO$_{2}$ are plotted in Fig.~\ref{Fig:ScatteringRateSiO2}.
Landau damping is taken into account by setting the coupling coefficient
of the IPP (SPP) modes to zero when $Q<Q_{c}$ ($Q\geq Q_{c}$). We
observe that at low energies, the IPP scattering rates are much higher
than the SPP ones. At higher energies, the SPP scattering rates increase
rapidly. The dominant scattering mechanism around the Fermi level
appears to be due to the plasmon-like branch III in SiO$_{2}$ and
HfO$_{2}$. In addition, at the Fermi level in $\mathrm{HfO_{2}}$,
the SPP branches have scattering rates comparable to those of branch
III. This explains why the low density mobility of HfO$_{2}$ is less
than that of SiO$_{2}$.

The expression for the IPP/SPP-limited part of the electrical conductivity
is: 
\begin{equation}
\sigma_{RP}=\frac{g_{s}g_{v}e^{2}}{4\pi\hbar^{2}k_{B}T}\int_{0}^{\infty}f(E-E_{F})[1-f(E-E_{F})]\Gamma_{tr}(E)^{-1}EdE\ .\label{Eq:ConductivityFormula}
\end{equation}
 where $g_{s}=2$ and $g_{v}=2$ are the spin and valley degeneracies
respectively. Only the contribution from the conduction band is included
in Eq.~(\ref{Eq:ConductivityFormula}). We use Eqs.~(\ref{Eq:MomentumRelaxationRate})
and (\ref{Eq:ConductivityFormula}) to compute the IPP/SPP-limited
electrical conductivity by setting: 
\begin{equation}
\Gamma_{tr}(E)=\Gamma_{RP}(s,\mathbf{K})\ .\label{Eq:SubLtdGamma}
\end{equation}
 In making this approximation, we ignore the other effects (ripples,
charged impurity, acoustic phonons, optical phonons, etc). The scattering
rates from the acoustic and optical phonons tend to be significantly
smaller and are not the limiting factor in electrical transport in
supported graphene~\cite{RSShishir:JPhys09}. Impurity scattering
tends to be the dominant limiting factor, but its effects can be reduced
by varying fabrication conditions. Thus, the conductivity using Eq.~(\ref{Eq:SubLtdGamma})
gives us its upper bound. We calculate the remote phonon-limited mobility
as: 
\begin{equation}
\mu_{RP}=\frac{\sigma_{RP}}{en}\label{Eq:SubLtdMobility}
\end{equation}
 where $n=\frac{g_{s}g_{v}}{2\pi\hbar^{2}v_{F}^{2}}\int_{0}^{\infty}f(E-E_{F})EdE$
is the carrier density. For $n=10^{12}\mathrm{cm^{-2}}$ in SiO$_{2}$,
we obtain $\mu_{RP}\approx40,000\ \mathrm{cm^{2}V^{-1}s^{-1}}$. This
is more than the corresponding values reported in the literature ($\sim1000-20,000\ \mathrm{cm^{2}V^{-1}s^{-1}}$)~\cite{YWTan:PRL07,KSNovoselov:Nature05}
although we have to bear in mind that it is an upper limit. Nonetheless,
it suggests that IPP/SPP scattering imposes a bound on the electron
mobility.

\subsection{Mobility results}

Although suspended graphene has an intrinsic mobility limit of $200,000\ \mathrm{cm^{2}V^{-1}s^{-1}}$
at room temperature~\cite{KIBolotin:SSC08}, typical numbers for
graphene on SiO$_{2}$ tend to fall in the range 1000-20,000 $\mathrm{cm^{2}V^{-1}s^{-1}}$~\cite{JHChen:NatureNanotech08}.
One significant reason for this drastic reduction in mobility is believed
to be the presence of charged impurities in the substrate which causes
long-range Coulombic scattering~\cite{SAdam:PNAS07,SAdam:SSC09,CJang:PRL08}
and much effort has been directed towards the amelioration of the
effects of these charged impurities. For example, it has been suggested
that modifying the dielectric environment of the graphene, either
through immersion in a high-$\kappa$ liquid or an overlayer of high-$\kappa$
dielectric material, can lead to a weakening of the Coulombic interaction
and an increase in electron mobility~\cite{CJang:PRL08}. On the
other hand, actual experimental evidence in favor of this theory is
ambiguous. Electrical conductivity data from Jang and co-workers~\cite{CJang:PRL08}
as well as Ponomarenko and co-workers~\cite{LAPonomarenko:PRL09}
indicate a smaller-than-expected increase in mobility when a liquid
overlayer is used. This suggests that mechanisms other than long and
short-range impurity scattering are at play here. Here, we turn to
the problem of scattering by IPP modes.

\subsubsection{Comparing different substrates}

Having set up the theoretical framework in the earlier sections, we
now apply it to the study of the remote phonon-limited mobility of
four commonly-used substrates: SiO$_{2}$, HfO$_{2}$, h-BN and Al$_{2}$O$_{3}$.
Silicon dioxide is the most common substrate material while HfO$_{2}$
and Al$_{2}$O$_{3}$ are high-$\kappa$ dielectrics commonly used
as top gate oxides~\cite{KZou:PRL10,NYGarces:JAP11}. Hexagonal boron
nitride shows much promise as both a substrate and a top gate dielectric
material~\cite{CRDean:NatureNanotech10}. The study of the remote
phonon-limited mobility in these substrates allows us to understand
how electronic transport in supported graphene depends on the frequencies
and relative permittivities of the substrate phonons.

From Eq.~(\ref{Eq:ConductivityFormula}) with the effects of Landau
damping taken into account, we compute the remote phonon-limited mobility
numerically, using the well-known Gilat-Raubenheimer method~\cite{GGilat:PR66}
to discretize the sum in Eq.~(\ref{Eq:ConductivityFormula}). We
plot $\mu_{RP}$ as a function of carrier density ($n=0.3\times10^{12}\mathrm{\ cm^{-2}}$
to $5.2\times10^{12}\mathrm{\ cm^{-2}}$) at 300 K in Fig.~\ref{Fig:ScreenedResults}.
Note that the mobility values of the high-$\kappa$ substrates (HfO$_{2}$
and Al$_{2}$O$_{3}$) are substantially lower compared to SiO$_{2}$
and h-BN in the carrier density range $n<2.0\times10^{12}\mathrm{cm^{-2}}$.
Similar results have been found in MOS systems~\cite{MVFischetti:JAP01}.
Hexagonal BN has the highest mobility at low carrier densities because
of its high phonon frequencies, which corresponds to low Bose-Einstein
occupancy, as well as its weak dipole coupling to graphene. In general,
$\mu_{RP}$ for all four substrates increases with $n$ because the
dynamic screening effect becomes stronger at higher carrier densities.
At low carrier densities, the mobility is low for all the substrates
because there is a large proportion of plasmons modes whose frequencies
are lower than the SPP mode frequencies. Thus, their coupling to the
SPP modes results in the formation of \emph{anti-screened} IPP modes
that couple \emph{more} strongly to the carriers, a phenomenon that
has been studied for polar semiconductors~\cite{BKRidley:Book99}.
However, as $n$ increases, the mobility for all four substrates rises
because the plasmon frequency scales as $\omega_{p}\propto n^{1/4}$,
resulting in higher-frequency plasmon modes. Thus, the plasmon-phonon
coupling forms screened IPP modes that are weakly coupled to the carriers.
Furthermore, at higher carrier densities, Landau damping becomes less
important as a result of the increasing magnitude of the plasmon wave
vector $Q_{c}$. Contrary to expectation, we find that the mobility
for HfO$_{2}$ exceeds those of other substrates at larger densities
($n=5\times10^{12}\mathrm{\ cm^{-2}}$). At $n=5\times10^{12}\mathrm{\ cm^{-2}}$,
HfO$_{2}$ has the highest remote-phonon mobility followed by h-BN,
SiO$_{2}$ and Al$_{2}$O$_{3}$. This is because the proportion of
screened IPP modes increases with increasing carrier density. Given
the small values of $\omega_{TO1}$ and $\omega_{TO2}$ for HfO$_{2}$,
its coupling coefficients are smaller as a result of stronger dynamic
screening. This weaker coupling compensates in part the higher occupation
factors. In contrast, the larger values of $\omega_{TO1}$ and $\omega_{TO2}$
for h-BN imply that screening does not play a significant role at
low carrier densities. Hence, its coupling to the graphene carriers
does not diminish as rapidly as carrier density increases. The computed
$\mu_{RP}$ values for $\mathrm{HfO_{2}}$ and h-BN highlight the
role of low-frequency excitations in carrier scattering. The low-frequency
modes are highly occupied at room temperature and induce carrier significant
scattering at low $n$. At higher $n$ when dynamic screening becomes
important, the low-frequency modes are more strongly screened and
their coupling to the carriers becomes diminished more rapidly than
that of high-frequency modes.

\subsubsection{Dynamic screening effects}

\begin{figure}
\includegraphics[width=4in]{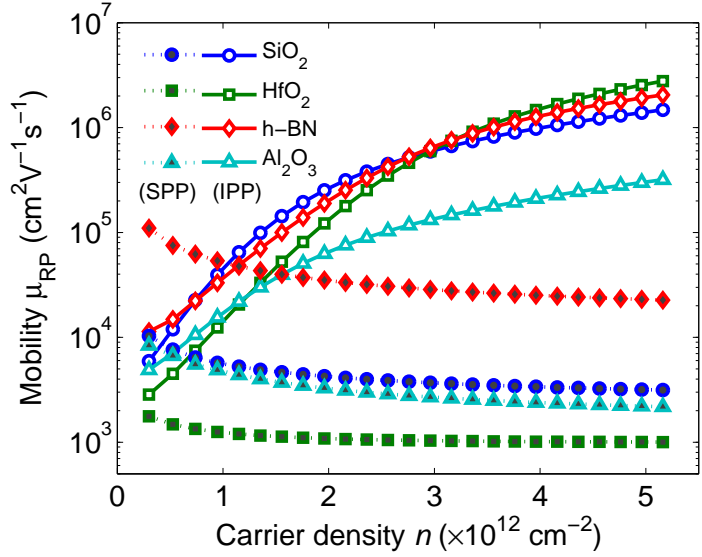} \caption{Calculated conductivity remote phonon-limited mobility for different
values of carrier density and different substrates (SiO$_{2}$, HfO$_{2}$,
h-BN and Al$_{2}$O$_{3}$) at room temperature (300 K). The IPP-limited
mobility values are plotted using solid lines with unfilled symbols
while the SPP-limited mobility values are plotted using dotted lines
with solid symbols.}

\label{Fig:ScreenedResults} 
\end{figure}

To compute the mobility for the case without any screening or anti-screening
effects, the Landau damping cutoff wave vector is decreased, \emph{i.e.},
$Q_{c}\rightarrow0$, resulting in the replacement of all the IPP
modes with SPP modes. We plot the SPP-limited mobility as a function
of carrier density in Fig.~\ref{Fig:ScreenedResults} (solid symbols),
and compare these results for the IPP-limited mobility. The SPP-limited
mobility for different substrates spans a range of values varying
over nearly two orders of magnitude. In the absence of dynamic screening
or anti-screening, the SPP-limited mobility for HfO$_{2}$ is only
around $\mathrm{1000\ cm^{2}V^{-1}s^{-1}}$ at $n=10^{12}\ \mathrm{cm}^{-2}$,
more than an order of magnitude smaller than the corresponding IPP-limited
mobility, because of its low phonon frequencies. This result is also
clearly inconsistent with experimental observations, since significantly
higher mobility values have been reported for HfO$_{2}$-covered graphene~\cite{BFallahazad:APL10,KZou:PRL10}.
The drastic reduction of the computed mobility suggests that screening
is very important for the determination of scattering rates in a coupled
plasmon-phonon system with low frequency modes. In contrast, h-BN
gives an SPP-limited mobility of $\sim110,000\ \mathrm{cm^{2}V^{-1}s^{-1}}$
at $n=0.3\times10^{12}\ \mathrm{cm}^{-2}$, which is still close to
the IPP-limited mobility, indicating that its high frequency modes
are relatively unaffected by screening. The maximum SPP-limited mobility
for Al$_{2}$O$_{3}$ is around $\mathrm{8,400\ cm^{2}V^{-1}s^{-1}}$at
$n=0.3\times10^{12}\ \mathrm{cm}^{-2}$, which is much smaller than
the $\mathrm{19000\ cm^{2}V^{-1}s^{-1}}$ extracted by Jandhyala and
co-workers~\cite{SJandhyala:ACSNano12} who used Al$_{2}$O$_{3}$
for their top gate dielectric. This disagreement reinforces the necessity
of including dynamic screening effects. Furthermore, the carrier density
dependence of SPP-limited mobility is different from that of IPP-limited
mobility. The IPP-limited $\mu_{RP}$ increases rapidly with carrier
density because dynamic screening becomes stronger at higher $n$,
an effect that is not found in SPP-limited mobility. In contrast,
SPP-limited $\mu_{RP}$ decreases monotonically with increasing $n$.

Our results suggest that HfO$_{2}$ remains a promising candidate
material for integration with graphene since its high static permittivity
can reduce the effect of charged impurities~\cite{AKonar:PRB10}
while its IPP scattering rates are relatively low when the carrier
density is high. Although its surface excitations are low-frequency,
which results in high Bose-Einstein occupancy, this is offset by its
relatively strong dynamic screening at higher carrier densities. Thus,
IPP scattering does not represent a problem for its integration with
graphene field-effect transistors. As expected, h-BN is also a good
dielectric material since its high phonon frequencies imply a low
Bose-Einstein occupation factor. Furthermore, its smooth interface
results in a smaller interface charge density and is less likely to
induce mobility-limiting ripples in graphene. 

\begin{table}
\begin{tabular}{|c|c|c|c|c|}
\hline 
 & $\mathrm{SiO_{2}}$  & h-BN  & $\mathrm{HfO_{2}}$  & $\mathrm{Al_{2}O_{3}}$\tabularnewline
\hline 
\hline 
$\epsilon_{ox}^{0}$ ($\epsilon_{0}$)  & 3.90  & 5.09  & 22.00  & 12.35\tabularnewline
\hline 
$\epsilon_{ox}^{i}$ ($\epsilon_{0}$)  & 3.05  & 4.57  & 6.58  & 7.27\tabularnewline
\hline 
$\epsilon_{ox}^{\infty}$ ($\epsilon_{0}$)  & 2.50  & 4.10  & 5.03  & 3.20\tabularnewline
\hline 
$\omega_{TO1}$ (meV)  & 55.60  & 97.40  & 12.40  & 48.18\tabularnewline
\hline 
$\omega_{TO2}$ (meV)  & 138.10  & 187.90  & 48.35  & 71.41\tabularnewline
\hline 
\end{tabular}\caption{Parameters {[}see Eq.~(\ref{Eq:DielectricEquation}){]} used in computing
dispersion relation and scattering rates for SiO$_{2}$, h-BN, HfO$_{2}$
and Al$_{2}$O$_{3}$. They are taken from Refs.~\cite{MVFischetti:JAP01}
and \cite{VPerebeinos:PRB10}.}

\label{Tab:SubstrateDielectricParameters} 
\end{table}

\subsubsection{Temperature dependence}

Remote phonon scattering exhibits a strong temperature dependence
-- stronger than for ionized impurity scattering -- because the Bose-Einstein
occupation of the remote phonons decreases with lower temperature.
This change in the distribution of the remote phonons (IPP or SPP)
necessarily implies that the electronic transport character of the
SLG must change with temperature. At lower temperatures, scattering
with the remote phonons decreases, resulting in a higher remote phonon-limited
electrical mobility. The dependence of the change in mobility with
temperature is related to the dispersion of the remote phonons and
their coupling to the graphene electrons. By measuring the dependence
of the mobility or conductivity with respect to temperature, it is
possible to determine the dominant scattering mechanisms in the supported
graphene. Given that our model of electron-IPP scattering differs
from the more common electron-SPP scattering model, comparing the
temperature dependence of the substrate-limited mobility can enable
us to distinguish between the two models.

\begin{figure}
\includegraphics[width=4in]{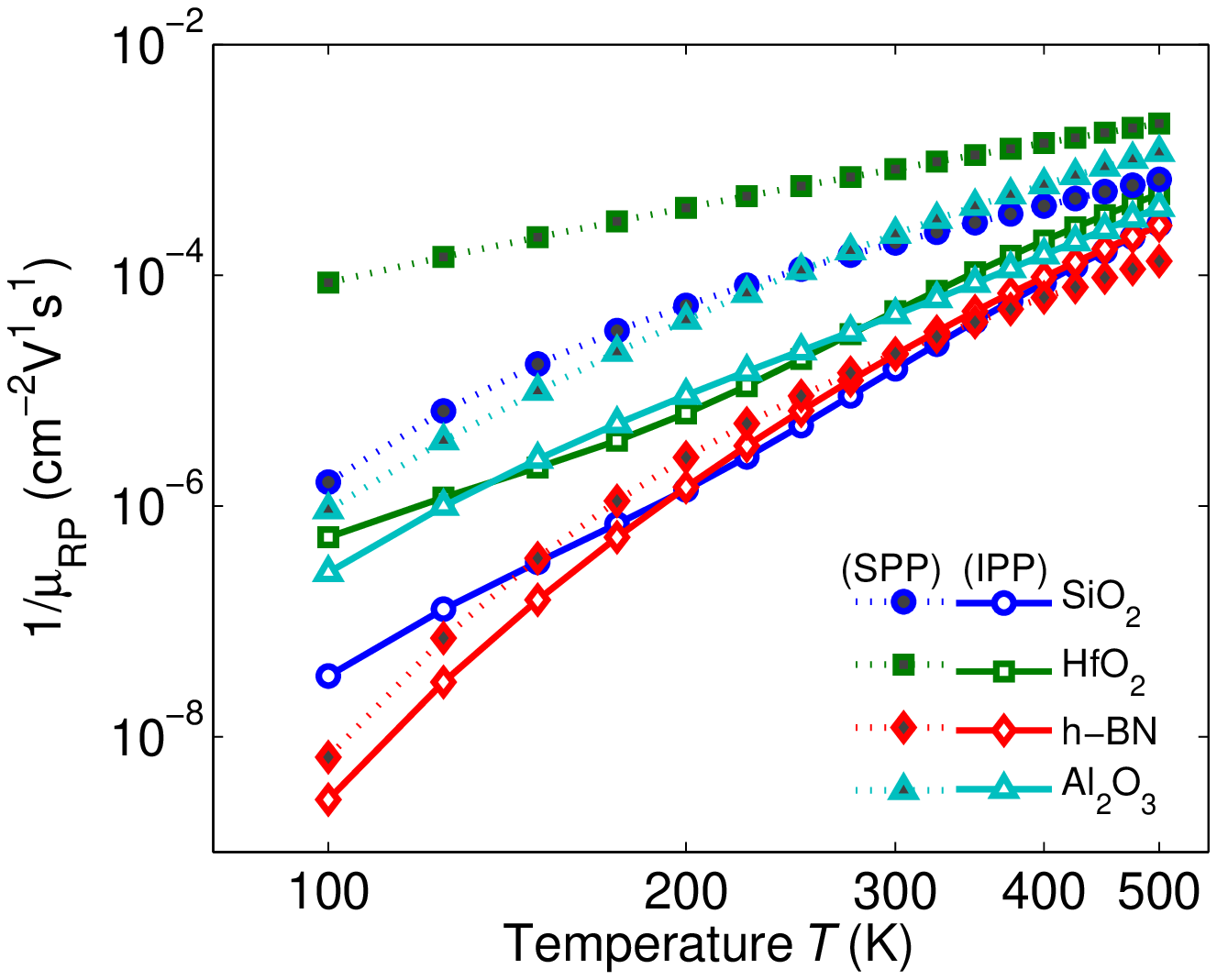}
\includegraphics[width=4in]{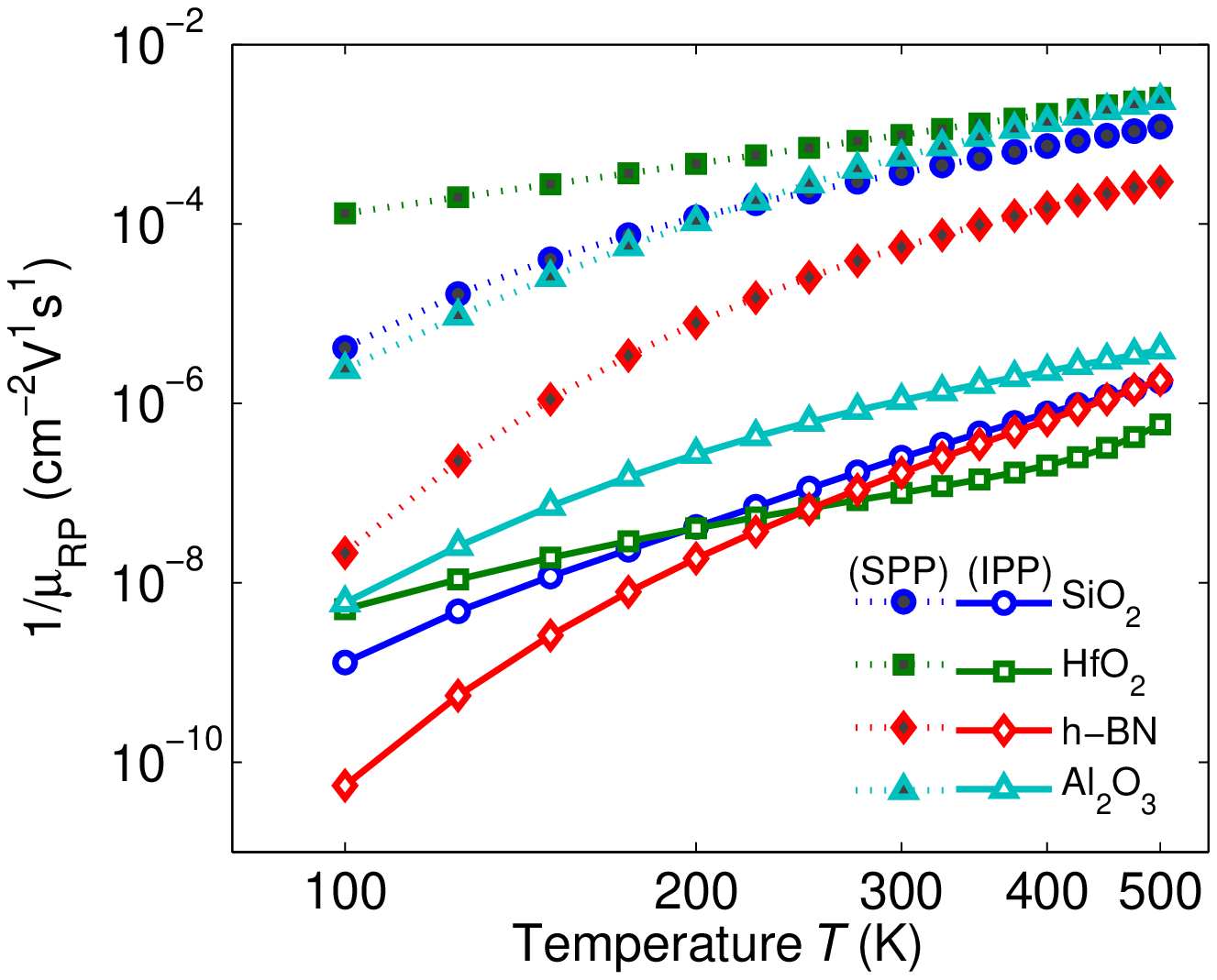}
\caption{Inverse SPP and IPP-limited mobility versus temperature at (a) $n=10^{12}\ \text{cm}^{-2}$
and (b) $n=10^{13}\ \text{cm}^{-2}$ for SiO$_{2}$, HfO$_{2}$, h-BN
and Al$_{2}$O$_{3}$. $1/\mu_{RP}$ is strongly temperature dependent
at low carrier densities only.}

\label{Fig:InverseMobility} 
\end{figure}

The mobility of supported graphene over the temperature range of 100
to 500 K for the different substrates is computed at carrier densities
of $n=10^{12}\ \text{cm}^{-2}$ and $n=10^{13}\ \text{cm}^{-2}$.
For the purpose of comparison, we perform the calculation for the
case with screening (IPP) and without screening (SPP). The results
($1/\mu_{RP}$ vs. $T$) are shown in Fig.~\ref{Fig:InverseMobility}a
and b. In Fig.~\ref{Fig:InverseMobility}a, we plot the IPP- and
SPP-limited inverse mobility at $n=10^{12}\ \text{cm}^{-2}$. As expected,
the substrate-limited mobility decreases with rising temperatures
for both the screened and unscreened cases. From the plots, we observe
that there exists an `activation' temperature for each substrate at
which the inverse mobility increases precipitously. For SiO$_{2}$,
that temperature is around 200 K in the screened case and around 120
K in the unscreened case. This difference is striking and may be used
to distinguish the IPP model from the SPP model at low carrier densities.
In all four substrates, the slope of $1/\mu_{RP}$ with respect to
$T$ is also steeper in the IPP-limited case than in the SPP-limited
case. In Fig.~\ref{Fig:InverseMobility}b, we plot again the IPP-
and SPP-limited inverse mobility but at a much higher carrier density
of $n=10^{13}\ \text{cm}^{-2}$. The IPP-limited $1/\mu_{RP}$ is
about three orders of magnitude smaller than the SPP-limited $1/\mu_{RP}$
from 100 to 500 K. At high carrier densities, IPP scattering is insignificant
and any changes in total mobility with respect to temperature cannot
attribute to IPP scattering.

The results in Fig.~\ref{Fig:InverseMobility} suggest that if IPP
modes are the surface excitations that limit carrier transport in
SiO$_{2}$-supported graphene at room temperature, then the mobility
would have a significant increase at around 200 K for $n=10^{12}\ \text{cm}^{-2}$.
However, this IPP temperature dependence disappears at much higher
carrier densities ($\sim n=10^{13}\ \text{cm}^{-2}$) because the
IPP coupling to electrons becomes so weak that it no longer contributes
significantly to carrier scattering. The results in Fig. \ref{Fig:InverseMobility}
also shows that the $\mu_{RP}$ increases monotonically with $n$.
This should be contrasted with the result of Fratini and Guinea \cite{SFratini:PRB08}
who found that $\mu_{RP}$ \emph{decreases} as $\sim1/\sqrt{n}$ at
room temperature. This is because the coupling coefficient $\lim_{Q\rightarrow\infty}M_{\mathbf{Q}}$,
which is proportional to the matrix element, scales as $1/\sqrt{Q}$
in the SPP model with static screening. In Fig. \ref{Fig:CouplingCoeffSiO2},
$\lim_{Q\rightarrow\infty}M_{\mathbf{Q}}Q$ decreases with $Q$, implying
that $\lim_{Q\rightarrow\infty}M_{\mathbf{Q}}$ scales as $Q^{\alpha}$
where $\alpha<-1$. In other words, the coupling coefficient vanishes
more rapidly with $Q$ in the IPP model than the SPP model. Our $\mu_{RP}$
results parallel those in Ref. \cite{ZRen:IEDM03} in which the remote
phonon-limited mobility increases with the carrier mobility in a 2-dimensional
electron gas system in the Si inversion layer with high-$\kappa$
insulators. 

In supported SLG, the carrier mobility is limited by three scattering
mechanisms: long-range charged impurity, short-range defect and remote
phonon scattering \cite{WZhu:PRB09}. The intrinsic phonon scattering
processes in graphene can be effectively neglected. Of the three scattering
mechanisms, only remote phonon scattering is strongly temperature
dependent. The IPP model suggests that remote phonon scattering diminishes
with increasing carrier density. Thus, the experimental consequence
is that the temperature dependence of the mobility in supported-SLG
should weaken at higher carrier densities. On the other hand, the
SPP model predicts that the temperature dependence of the mobility
should increase at higher carrier densities \cite{SFratini:PRB08}.
This difference in the temperature dependence of the total mobility
between the two models should be easily discriminable in experiments.

\subsubsection{Disordered graphene}

We discuss qualitatively the interfacial plasmon-phonon phenomenon
in disordered graphene. It is well-known that graphene grown by chemical
vapor deposition (CVD) \cite{XLi:Science09} is generally polycrystalline
and contains a high density of defects. In supported graphene, charged
impurities from the substrate and other defects scatter graphene carriers.
These defects can affect the dynamics of plasmons in graphene which
may in turn affect the hybridization between the plasmon and the SPP
modes. At short wavelengths, the plasmon lifetime rapidly decreases
as a result of Landau damping which results in the decay of the plasmons
into single-particle excitations. At long wavelengths, the plasmon
lifetime can be affected by defects in the graphene. As far as we
know, there is no theory of graphene plasmon damping from defects.
However, it has been pointed out that long-wavelength plasmons in
polycrystalline metal undergo anomalously large damping due to scattering
with structural defects \cite{VKrishan:PRL70}. If this is also true
in polycrystalline or defective graphene, then it implies that the
long-wavelength surface excitation in supported graphene are SPP,
not IPP, modes. 

\begin{figure}
\includegraphics[width=4in]{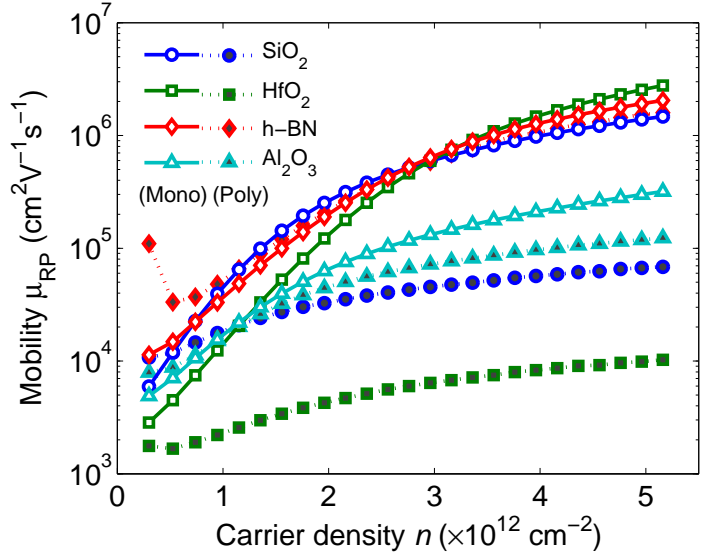}
\caption{Remote phonon-limited mobility in perfect monocrystalline (clear symbols)
and defective polycrystalline graphene (solid symbols) for SiO$_{2}$,
HfO$_{2}$, h-BN and Al$_{2}$O$_{3}$. As carrier density increases,
$\mu_{RP}$ also increases. The use of long-wavelength SPP modes leads
to a significant decrease in $\mu_{RP}$ in polycrystalline graphene.}

\label{Fig:DefectGrapheneMobility} 
\end{figure}

To model phenomenologically this damping of long-wavelength plasmon
modes in polycrystalline graphene with defects, we set another cutoff
wave vector $Q_{d}$ below which the surface excitations are SPP and
not IPP modes. $Q_{d}$ is possibly related to the length scale $\lambda$
of the inhomogeneities or defects in graphene. As a guess, we choose
$\lambda$= 6 nm, which is a typical autocorrelation length of `puddles'
in neutral supported graphene\cite{SAdam:PRB11}, and set $Q_{d}=1/\lambda$.
Hence, in our model, for $Q\geq Q_{c}$ and $Q\leq Q_{d}$, the surface
excitations are SPP modes while for $Q_{d}<Q<Q_{c}$, they are IPP
modes. We compute the remote phonon-limited mobility at 300 K and
plot the results in Fig. \ref{Fig:DefectGrapheneMobility}. We find
that the long-wavelength SPP dramatically alters the carrier dependence
of $\mu_{RP}$ in SiO$_{2}$, HfO$_{2}$ and Al$_{2}$O$_{3}$. In
perfect monocrystalline graphene, $\mu_{RP}$ reaches $\mathrm{\sim2\times10^{6}\ cm^{2}V^{-1}s^{-1}}$
in HfO$_{2}$ and SiO$_{2}$ at $n=5\times10^{12}\ \text{cm}^{-2}$
. On the other hand, in polycrystalline graphene with defects, it
drops to the range of $10^{4}$ to $\mathrm{10^{5}\ cm^{2}V^{-1}s^{-1}}$.
For h-BN, $\mu_{RP}$ is quite relatively unaffected by the long-wavelength
SPP modes except at low carrier densities ($n<0.5\times10^{6}\ cm^{-2}$). 

This change in remote phonon-limited mobility highlights the possible
role of defects in the surface excitations of supported graphene.
We emphasize that our treatment is purely phenomenological and a more
rigorous treatment of plasmon damping is needed in order to obtain
a more quantitatively accurate model. Nevertheless, it emphasizes
the relationship between dynamic screening and plasmons. In highly
defective graphene, the surface excitations may be unscreened SPPs
rather than IPPs because of plasmon damping. This should be taken
into account when interpreting electronic transport experimental data
of exfoliated and CVD-grown graphene.

\section{Conclusion}

We have studied coupled interfacial plasmon-phonon excitations in
supported graphene. The coupling between the pure graphene plasmon
and the surface polar phonon modes of the substrates results in the
formation of the IPP modes, and this coupling is responsible for the
screening and anti-screening of the IPP modes. Accounting for these
modes, we calculate the room temperature scattering rates and substate-limited
mobility for $\mathrm{SiO_{2}}$, $\mathrm{HfO_{2}}$, h-BN and $\mathrm{Al_{2}O_{3}}$
at different carrier densities. The results suggest that, despite
being a high-$\kappa$ oxide with low frequency modes, $\mathrm{HfO_{2}}$
exhibits a substrate-limited mobility comparable to that of h-BN at
high carrier densities. We attribute this to the dynamic screening
of the $\mathrm{HfO_{2}}$ low-frequency modes. The disadvantage of
the higher Bose-Einstein occupation of these low-frequency modes is
offset by the stronger dynamic screening which suppresses the electron-IPP
coupling. Our study also indicates that the contribution to scattering
by high-frequency substrate phonon modes cannot be neglected because
of they are less weakly screened by the graphene plasmons. The temperature
dependence of the remote phonon-limited mobility is also calculated
within out theory. Its change with temperature is different at low
and high carrier densities. We find that in the IPP model, the temperature
dependence of the mobility diminishes with increasing carrier density
only, in direct contrast to the predictions of the more commonly used
SPP models. The implications of the damping of long-wavelength plasmons
have also been studied. We find that the it leads to a substantial
reduction in the remote phonon-limited mobility in SiO$_{2}$, HfO$_{2}$
and Al$_{2}$O$_{3}$. 

We gratefully acknowledge the support provided by Texas Instruments,
the Semiconductor Research Corporation (SRC), the Microelectronics
Advanced Research Corporation (MARCO), the Focus Center Research Project
(FCRP) for Materials, Structures and Devices (MSD), and Samsung Electronics
Ltd. We also like to thank David K. Ferry (Arizona State University),
Eric Pop (University of Illinois), and Andrey Serov (University of
Illinois) for engaging in valuable technical discussions.

\bibliographystyle{apsrev4-1}
\bibliography{SubstrateRemotePhonons_v23July2012}

\begin{thebibliography}{52}%
\makeatletter
\providecommand \@ifxundefined [1]{%
 \@ifx{#1\undefined}
}%
\providecommand \@ifnum [1]{%
 \ifnum #1\expandafter \@firstoftwo
 \else \expandafter \@secondoftwo
 \fi
}%
\providecommand \@ifx [1]{%
 \ifx #1\expandafter \@firstoftwo
 \else \expandafter \@secondoftwo
 \fi
}%
\providecommand \natexlab [1]{#1}%
\providecommand \enquote  [1]{``#1''}%
\providecommand \bibnamefont  [1]{#1}%
\providecommand \bibfnamefont [1]{#1}%
\providecommand \citenamefont [1]{#1}%
\providecommand \href@noop [0]{\@secondoftwo}%
\providecommand \href [0]{\begingroup \@sanitize@url \@href}%
\providecommand \@href[1]{\@@startlink{#1}\@@href}%
\providecommand \@@href[1]{\endgroup#1\@@endlink}%
\providecommand \@sanitize@url [0]{\catcode `\\12\catcode `\$12\catcode
  `\&12\catcode `\#12\catcode `\^12\catcode `\_12\catcode `\%12\relax}%
\providecommand \@@startlink[1]{}%
\providecommand \@@endlink[0]{}%
\providecommand \url  [0]{\begingroup\@sanitize@url \@url }%
\providecommand \@url [1]{\endgroup\@href {#1}{\urlprefix }}%
\providecommand \urlprefix  [0]{URL }%
\providecommand \Eprint [0]{\href }%
\providecommand \doibase [0]{http://dx.doi.org/}%
\providecommand \selectlanguage [0]{\@gobble}%
\providecommand \bibinfo  [0]{\@secondoftwo}%
\providecommand \bibfield  [0]{\@secondoftwo}%
\providecommand \translation [1]{[#1]}%
\providecommand \BibitemOpen [0]{}%
\providecommand \bibitemStop [0]{}%
\providecommand \bibitemNoStop [0]{.\EOS\space}%
\providecommand \EOS [0]{\spacefactor3000\relax}%
\providecommand \BibitemShut  [1]{\csname bibitem#1\endcsname}%
\let\auto@bib@innerbib\@empty
\bibitem [{\citenamefont {Novoselov}\ \emph {et~al.}(2005)\citenamefont
  {Novoselov}, \citenamefont {Geim}, \citenamefont {Morozov}, \citenamefont
  {Jiang}, \citenamefont {Grigorieva}, \citenamefont {Dubonos},\ and\
  \citenamefont {Firsov}}]{KSNovoselov:Nature05}%
  \BibitemOpen
  \bibfield  {author} {\bibinfo {author} {\bibfnamefont {K.}~\bibnamefont
  {Novoselov}}, \bibinfo {author} {\bibfnamefont {A.}~\bibnamefont {Geim}},
  \bibinfo {author} {\bibfnamefont {S.}~\bibnamefont {Morozov}}, \bibinfo
  {author} {\bibfnamefont {D.}~\bibnamefont {Jiang}}, \bibinfo {author}
  {\bibfnamefont {M.}~\bibnamefont {Grigorieva}}, \bibinfo {author}
  {\bibfnamefont {S.}~\bibnamefont {Dubonos}}, \ and\ \bibinfo {author}
  {\bibfnamefont {A.}~\bibnamefont {Firsov}},\ }\href@noop {} {\bibfield
  {journal} {\bibinfo  {journal} {Nature}\ }\textbf {\bibinfo {volume} {438}},\
  \bibinfo {pages} {197} (\bibinfo {year} {2005})}\BibitemShut {NoStop}%
\bibitem [{\citenamefont {Hwang}\ \emph {et~al.}(2007)\citenamefont {Hwang},
  \citenamefont {Adam},\ and\ \citenamefont {Das~Sarma}}]{EHHwang:PRL07}%
  \BibitemOpen
  \bibfield  {author} {\bibinfo {author} {\bibfnamefont {E.~H.}\ \bibnamefont
  {Hwang}}, \bibinfo {author} {\bibfnamefont {S.}~\bibnamefont {Adam}}, \ and\
  \bibinfo {author} {\bibfnamefont {S.}~\bibnamefont {Das~Sarma}},\ }\href@noop
  {} {\bibfield  {journal} {\bibinfo  {journal} {Phys. Rev. Lett.}\ }\textbf
  {\bibinfo {volume} {98}},\ \bibinfo {pages} {186806} (\bibinfo {year}
  {2007})}\BibitemShut {NoStop}%
\bibitem [{\citenamefont {Balandin}\ \emph {et~al.}(2008)\citenamefont
  {Balandin}, \citenamefont {Ghosh}, \citenamefont {Bao}, \citenamefont
  {Calizo}, \citenamefont {Teweldebrhan}, \citenamefont {Miao},\ and\
  \citenamefont {Lau}}]{AABalandin:NL08}%
  \BibitemOpen
  \bibfield  {author} {\bibinfo {author} {\bibfnamefont {A.}~\bibnamefont
  {Balandin}}, \bibinfo {author} {\bibfnamefont {S.}~\bibnamefont {Ghosh}},
  \bibinfo {author} {\bibfnamefont {W.}~\bibnamefont {Bao}}, \bibinfo {author}
  {\bibfnamefont {I.}~\bibnamefont {Calizo}}, \bibinfo {author} {\bibfnamefont
  {D.}~\bibnamefont {Teweldebrhan}}, \bibinfo {author} {\bibfnamefont
  {F.}~\bibnamefont {Miao}}, \ and\ \bibinfo {author} {\bibfnamefont
  {C.}~\bibnamefont {Lau}},\ }\href@noop {} {\bibfield  {journal} {\bibinfo
  {journal} {Nano Lett.}\ }\textbf {\bibinfo {volume} {8}},\ \bibinfo {pages}
  {902} (\bibinfo {year} {2008})}\BibitemShut {NoStop}%
\bibitem [{\citenamefont {Bolotin}\ \emph {et~al.}(2008)\citenamefont
  {Bolotin}, \citenamefont {Sikes}, \citenamefont {Jiang}, \citenamefont
  {Klima}, \citenamefont {Fudenberg}, \citenamefont {Hone}, \citenamefont
  {Kim},\ and\ \citenamefont {Stormer}}]{KIBolotin:SSC08}%
  \BibitemOpen
  \bibfield  {author} {\bibinfo {author} {\bibfnamefont {K.}~\bibnamefont
  {Bolotin}}, \bibinfo {author} {\bibfnamefont {K.}~\bibnamefont {Sikes}},
  \bibinfo {author} {\bibfnamefont {Z.}~\bibnamefont {Jiang}}, \bibinfo
  {author} {\bibfnamefont {M.}~\bibnamefont {Klima}}, \bibinfo {author}
  {\bibfnamefont {G.}~\bibnamefont {Fudenberg}}, \bibinfo {author}
  {\bibfnamefont {J.}~\bibnamefont {Hone}}, \bibinfo {author} {\bibfnamefont
  {P.}~\bibnamefont {Kim}}, \ and\ \bibinfo {author} {\bibfnamefont
  {H.}~\bibnamefont {Stormer}},\ }\href@noop {} {\bibfield  {journal} {\bibinfo
   {journal} {Solid State Commun.}\ }\textbf {\bibinfo {volume} {146}},\
  \bibinfo {pages} {351} (\bibinfo {year} {2008})}\BibitemShut {NoStop}%
\bibitem [{\citenamefont {Lemme}\ \emph {et~al.}(2008)\citenamefont {Lemme},
  \citenamefont {Echtermeyer}, \citenamefont {Baus}, \citenamefont {Szafranek},
  \citenamefont {Bolten}, \citenamefont {Schmidt}, \citenamefont {Wahlbrink},\
  and\ \citenamefont {Kurz}}]{MCLemme:SSE08}%
  \BibitemOpen
  \bibfield  {author} {\bibinfo {author} {\bibfnamefont {M.}~\bibnamefont
  {Lemme}}, \bibinfo {author} {\bibfnamefont {T.}~\bibnamefont {Echtermeyer}},
  \bibinfo {author} {\bibfnamefont {M.}~\bibnamefont {Baus}}, \bibinfo {author}
  {\bibfnamefont {B.}~\bibnamefont {Szafranek}}, \bibinfo {author}
  {\bibfnamefont {J.}~\bibnamefont {Bolten}}, \bibinfo {author} {\bibfnamefont
  {M.}~\bibnamefont {Schmidt}}, \bibinfo {author} {\bibfnamefont
  {T.}~\bibnamefont {Wahlbrink}}, \ and\ \bibinfo {author} {\bibfnamefont
  {H.}~\bibnamefont {Kurz}},\ }\href@noop {} {\bibfield  {journal} {\bibinfo
  {journal} {Solid-State Electron.}\ }\textbf {\bibinfo {volume} {52}},\
  \bibinfo {pages} {514} (\bibinfo {year} {2008})}\BibitemShut {NoStop}%
\bibitem [{\citenamefont {Moon}\ \emph {et~al.}(2010)\citenamefont {Moon},
  \citenamefont {Curtis}, \citenamefont {Bui}, \citenamefont {Hu},
  \citenamefont {Gaskill}, \citenamefont {Tedesco}, \citenamefont {Asbeck},
  \citenamefont {Jernigan}, \citenamefont {VanMil}, \citenamefont {Myers-Ward}
  \emph {et~al.}}]{JSMoon:EDL10}%
  \BibitemOpen
  \bibfield  {author} {\bibinfo {author} {\bibfnamefont {J.}~\bibnamefont
  {Moon}}, \bibinfo {author} {\bibfnamefont {D.}~\bibnamefont {Curtis}},
  \bibinfo {author} {\bibfnamefont {S.}~\bibnamefont {Bui}}, \bibinfo {author}
  {\bibfnamefont {M.}~\bibnamefont {Hu}}, \bibinfo {author} {\bibfnamefont
  {D.}~\bibnamefont {Gaskill}}, \bibinfo {author} {\bibfnamefont
  {J.}~\bibnamefont {Tedesco}}, \bibinfo {author} {\bibfnamefont
  {P.}~\bibnamefont {Asbeck}}, \bibinfo {author} {\bibfnamefont
  {G.}~\bibnamefont {Jernigan}}, \bibinfo {author} {\bibfnamefont
  {B.}~\bibnamefont {VanMil}}, \bibinfo {author} {\bibfnamefont
  {R.}~\bibnamefont {Myers-Ward}},  \emph {et~al.},\ }\href@noop {} {\bibfield
  {journal} {\bibinfo  {journal} {Electron Device Letters, IEEE}\ }\textbf
  {\bibinfo {volume} {31}},\ \bibinfo {pages} {260} (\bibinfo {year}
  {2010})}\BibitemShut {NoStop}%
\bibitem [{\citenamefont {Pezoldt}\ \emph {et~al.}(2010)\citenamefont
  {Pezoldt}, \citenamefont {Hummel}, \citenamefont {Hanisch}, \citenamefont
  {Hotovy}, \citenamefont {Kadlecikova},\ and\ \citenamefont
  {Schwierz}}]{JPezoldt:PSS10}%
  \BibitemOpen
  \bibfield  {author} {\bibinfo {author} {\bibfnamefont {J.}~\bibnamefont
  {Pezoldt}}, \bibinfo {author} {\bibfnamefont {C.}~\bibnamefont {Hummel}},
  \bibinfo {author} {\bibfnamefont {A.}~\bibnamefont {Hanisch}}, \bibinfo
  {author} {\bibfnamefont {I.}~\bibnamefont {Hotovy}}, \bibinfo {author}
  {\bibfnamefont {M.}~\bibnamefont {Kadlecikova}}, \ and\ \bibinfo {author}
  {\bibfnamefont {F.}~\bibnamefont {Schwierz}},\ }\href@noop {} {\bibfield
  {journal} {\bibinfo  {journal} {Phys. Status Solidi C}\ }\textbf {\bibinfo
  {volume} {7}},\ \bibinfo {pages} {390} (\bibinfo {year} {2010})}\BibitemShut
  {NoStop}%
\bibitem [{\citenamefont {Fischetti}\ \emph {et~al.}(2001)\citenamefont
  {Fischetti}, \citenamefont {Neumayer},\ and\ \citenamefont
  {Cartier}}]{MVFischetti:JAP01}%
  \BibitemOpen
  \bibfield  {author} {\bibinfo {author} {\bibfnamefont {M.~V.}\ \bibnamefont
  {Fischetti}}, \bibinfo {author} {\bibfnamefont {D.~A.}\ \bibnamefont
  {Neumayer}}, \ and\ \bibinfo {author} {\bibfnamefont {E.~A.}\ \bibnamefont
  {Cartier}},\ }\href@noop {} {\bibfield  {journal} {\bibinfo  {journal} {J.
  Appl. Phys.}\ }\textbf {\bibinfo {volume} {90}},\ \bibinfo {pages} {4587}
  (\bibinfo {year} {2001})}\BibitemShut {NoStop}%
\bibitem [{\citenamefont {Fuchs}\ and\ \citenamefont
  {Kliewer}(1965)}]{RFuchs:PR65}%
  \BibitemOpen
  \bibfield  {author} {\bibinfo {author} {\bibfnamefont {R.}~\bibnamefont
  {Fuchs}}\ and\ \bibinfo {author} {\bibfnamefont {K.~L.}\ \bibnamefont
  {Kliewer}},\ }\href@noop {} {\bibfield  {journal} {\bibinfo  {journal} {Phys.
  Rev.}\ }\textbf {\bibinfo {volume} {140}},\ \bibinfo {pages} {A2076}
  (\bibinfo {year} {1965})}\BibitemShut {NoStop}%
\bibitem [{\citenamefont {Hess}\ and\ \citenamefont
  {Vogl}(1979)}]{KHess:SSC79}%
  \BibitemOpen
  \bibfield  {author} {\bibinfo {author} {\bibfnamefont {K.}~\bibnamefont
  {Hess}}\ and\ \bibinfo {author} {\bibfnamefont {P.}~\bibnamefont {Vogl}},\
  }\href@noop {} {\bibfield  {journal} {\bibinfo  {journal} {Solid State
  Commun.}\ }\textbf {\bibinfo {volume} {30}},\ \bibinfo {pages} {797}
  (\bibinfo {year} {1979})}\BibitemShut {NoStop}%
\bibitem [{\citenamefont {O'Regan}\ and\ \citenamefont
  {Fischetti}(2007)}]{TORegan:JCompElec07}%
  \BibitemOpen
  \bibfield  {author} {\bibinfo {author} {\bibfnamefont {T.}~\bibnamefont
  {O'Regan}}\ and\ \bibinfo {author} {\bibfnamefont {M.}~\bibnamefont
  {Fischetti}},\ }\href@noop {} {\bibfield  {journal} {\bibinfo  {journal} {J.
  Comput. Electron.}\ }\textbf {\bibinfo {volume} {6}},\ \bibinfo {pages} {81}
  (\bibinfo {year} {2007})}\BibitemShut {NoStop}%
\bibitem [{\citenamefont {Xiu}(2011)}]{KXiu:SISPAD11}%
  \BibitemOpen
  \bibfield  {author} {\bibinfo {author} {\bibfnamefont {K.}~\bibnamefont
  {Xiu}},\ }in\ \href@noop {} {\emph {\bibinfo {booktitle} {Simulation of
  Semiconductor Processes and Devices (SISPAD), 2011 International Conference
  on}}}\ (\bibinfo {organization} {IEEE},\ \bibinfo {year} {2011})\ pp.\
  \bibinfo {pages} {35--38}\BibitemShut {NoStop}%
\bibitem [{\citenamefont {Dean}\ \emph {et~al.}(2010)\citenamefont {Dean},
  \citenamefont {Young}, \citenamefont {Meric}, \citenamefont {Lee},
  \citenamefont {Wang}, \citenamefont {Sorgenfrei}, \citenamefont {Watanabe},
  \citenamefont {Taniguchi}, \citenamefont {Kim}, \citenamefont {Shepard} \emph
  {et~al.}}]{CRDean:NatureNanotech10}%
  \BibitemOpen
  \bibfield  {author} {\bibinfo {author} {\bibfnamefont {C.}~\bibnamefont
  {Dean}}, \bibinfo {author} {\bibfnamefont {A.}~\bibnamefont {Young}},
  \bibinfo {author} {\bibfnamefont {I.}~\bibnamefont {Meric}}, \bibinfo
  {author} {\bibfnamefont {C.}~\bibnamefont {Lee}}, \bibinfo {author}
  {\bibfnamefont {L.}~\bibnamefont {Wang}}, \bibinfo {author} {\bibfnamefont
  {S.}~\bibnamefont {Sorgenfrei}}, \bibinfo {author} {\bibfnamefont
  {K.}~\bibnamefont {Watanabe}}, \bibinfo {author} {\bibfnamefont
  {T.}~\bibnamefont {Taniguchi}}, \bibinfo {author} {\bibfnamefont
  {P.}~\bibnamefont {Kim}}, \bibinfo {author} {\bibfnamefont {K.}~\bibnamefont
  {Shepard}},  \emph {et~al.},\ }\href@noop {} {\bibfield  {journal} {\bibinfo
  {journal} {Nature Nanotechnology}\ }\textbf {\bibinfo {volume} {5}},\
  \bibinfo {pages} {722} (\bibinfo {year} {2010})}\BibitemShut {NoStop}%
\bibitem [{\citenamefont {Chen}\ \emph {et~al.}(2008)\citenamefont {Chen},
  \citenamefont {Jang}, \citenamefont {Xiao}, \citenamefont {Ishigami},\ and\
  \citenamefont {Fuhrer}}]{JHChen:NatureNanotech08}%
  \BibitemOpen
  \bibfield  {author} {\bibinfo {author} {\bibfnamefont {J.}~\bibnamefont
  {Chen}}, \bibinfo {author} {\bibfnamefont {C.}~\bibnamefont {Jang}}, \bibinfo
  {author} {\bibfnamefont {S.}~\bibnamefont {Xiao}}, \bibinfo {author}
  {\bibfnamefont {M.}~\bibnamefont {Ishigami}}, \ and\ \bibinfo {author}
  {\bibfnamefont {M.}~\bibnamefont {Fuhrer}},\ }\href@noop {} {\bibfield
  {journal} {\bibinfo  {journal} {Nature Nanotechnology}\ }\textbf {\bibinfo
  {volume} {3}},\ \bibinfo {pages} {206} (\bibinfo {year} {2008})}\BibitemShut
  {NoStop}%
\bibitem [{\citenamefont {Dorgan}\ \emph {et~al.}(2010)\citenamefont {Dorgan},
  \citenamefont {Bae},\ and\ \citenamefont {Pop}}]{VEDorgan:APL10}%
  \BibitemOpen
  \bibfield  {author} {\bibinfo {author} {\bibfnamefont {V.~E.}\ \bibnamefont
  {Dorgan}}, \bibinfo {author} {\bibfnamefont {M.-H.}\ \bibnamefont {Bae}}, \
  and\ \bibinfo {author} {\bibfnamefont {E.}~\bibnamefont {Pop}},\ }\href@noop
  {} {\bibfield  {journal} {\bibinfo  {journal} {Appl. Phys. Lett.}\ }\textbf
  {\bibinfo {volume} {97}},\ \bibinfo {pages} {082112} (\bibinfo {year}
  {2010})}\BibitemShut {NoStop}%
\bibitem [{\citenamefont {Robinson}\ \emph {et~al.}(2009)\citenamefont
  {Robinson}, \citenamefont {Wetherington}, \citenamefont {Tedesco},
  \citenamefont {Campbell}, \citenamefont {Weng}, \citenamefont {Stitt},
  \citenamefont {Fanton}, \citenamefont {Frantz}, \citenamefont {Snyder},
  \citenamefont {VanMil} \emph {et~al.}}]{JARobinson:NL09}%
  \BibitemOpen
  \bibfield  {author} {\bibinfo {author} {\bibfnamefont {J.}~\bibnamefont
  {Robinson}}, \bibinfo {author} {\bibfnamefont {M.}~\bibnamefont
  {Wetherington}}, \bibinfo {author} {\bibfnamefont {J.}~\bibnamefont
  {Tedesco}}, \bibinfo {author} {\bibfnamefont {P.}~\bibnamefont {Campbell}},
  \bibinfo {author} {\bibfnamefont {X.}~\bibnamefont {Weng}}, \bibinfo {author}
  {\bibfnamefont {J.}~\bibnamefont {Stitt}}, \bibinfo {author} {\bibfnamefont
  {M.}~\bibnamefont {Fanton}}, \bibinfo {author} {\bibfnamefont
  {E.}~\bibnamefont {Frantz}}, \bibinfo {author} {\bibfnamefont
  {D.}~\bibnamefont {Snyder}}, \bibinfo {author} {\bibfnamefont
  {B.}~\bibnamefont {VanMil}},  \emph {et~al.},\ }\href@noop {} {\bibfield
  {journal} {\bibinfo  {journal} {Nano Lett.}\ }\textbf {\bibinfo {volume}
  {9}},\ \bibinfo {pages} {2873} (\bibinfo {year} {2009})}\BibitemShut
  {NoStop}%
\bibitem [{\citenamefont {Sutter}(2009)}]{PSutter:NatureMat09}%
  \BibitemOpen
  \bibfield  {author} {\bibinfo {author} {\bibfnamefont {P.}~\bibnamefont
  {Sutter}},\ }\href@noop {} {\bibfield  {journal} {\bibinfo  {journal} {Nature
  Materials}\ }\textbf {\bibinfo {volume} {8}},\ \bibinfo {pages} {171}
  (\bibinfo {year} {2009})}\BibitemShut {NoStop}%
\bibitem [{\citenamefont {Fratini}\ and\ \citenamefont
  {Guinea}(2008)}]{SFratini:PRB08}%
  \BibitemOpen
  \bibfield  {author} {\bibinfo {author} {\bibfnamefont {S.}~\bibnamefont
  {Fratini}}\ and\ \bibinfo {author} {\bibfnamefont {F.}~\bibnamefont
  {Guinea}},\ }\href@noop {} {\bibfield  {journal} {\bibinfo  {journal} {Phys.
  Rev. B}\ }\textbf {\bibinfo {volume} {77}},\ \bibinfo {pages} {195415}
  (\bibinfo {year} {2008})}\BibitemShut {NoStop}%
\bibitem [{\citenamefont {Rotkin}\ \emph {et~al.}(2009)\citenamefont {Rotkin},
  \citenamefont {Perebeinos}, \citenamefont {Petrov},\ and\ \citenamefont
  {Avouris}}]{SVRotkin:NL09}%
  \BibitemOpen
  \bibfield  {author} {\bibinfo {author} {\bibfnamefont {S.}~\bibnamefont
  {Rotkin}}, \bibinfo {author} {\bibfnamefont {V.}~\bibnamefont {Perebeinos}},
  \bibinfo {author} {\bibfnamefont {A.}~\bibnamefont {Petrov}}, \ and\ \bibinfo
  {author} {\bibfnamefont {P.}~\bibnamefont {Avouris}},\ }\href@noop {}
  {\bibfield  {journal} {\bibinfo  {journal} {Nano Lett.}\ }\textbf {\bibinfo
  {volume} {9}},\ \bibinfo {pages} {1850} (\bibinfo {year} {2009})}\BibitemShut
  {NoStop}%
\bibitem [{\citenamefont {Perebeinos}\ \emph {et~al.}(2008)\citenamefont
  {Perebeinos}, \citenamefont {Rotkin}, \citenamefont {Petrov},\ and\
  \citenamefont {Avouris}}]{VPerebeinos:NL08}%
  \BibitemOpen
  \bibfield  {author} {\bibinfo {author} {\bibfnamefont {V.}~\bibnamefont
  {Perebeinos}}, \bibinfo {author} {\bibfnamefont {S.}~\bibnamefont {Rotkin}},
  \bibinfo {author} {\bibfnamefont {A.}~\bibnamefont {Petrov}}, \ and\ \bibinfo
  {author} {\bibfnamefont {P.}~\bibnamefont {Avouris}},\ }\href@noop {}
  {\bibfield  {journal} {\bibinfo  {journal} {Nano Lett.}\ }\textbf {\bibinfo
  {volume} {9}},\ \bibinfo {pages} {312} (\bibinfo {year} {2008})}\BibitemShut
  {NoStop}%
\bibitem [{\citenamefont {Konar}\ \emph {et~al.}(2010)\citenamefont {Konar},
  \citenamefont {Fang},\ and\ \citenamefont {Jena}}]{AKonar:PRB10}%
  \BibitemOpen
  \bibfield  {author} {\bibinfo {author} {\bibfnamefont {A.}~\bibnamefont
  {Konar}}, \bibinfo {author} {\bibfnamefont {T.}~\bibnamefont {Fang}}, \ and\
  \bibinfo {author} {\bibfnamefont {D.}~\bibnamefont {Jena}},\ }\href@noop {}
  {\bibfield  {journal} {\bibinfo  {journal} {Phys. Rev. B}\ }\textbf {\bibinfo
  {volume} {82}},\ \bibinfo {pages} {115452} (\bibinfo {year}
  {2010})}\BibitemShut {NoStop}%
\bibitem [{\citenamefont {Viljas}\ and\ \citenamefont
  {Heikkil{\"a}}(2010)}]{JKViljas:PRB10}%
  \BibitemOpen
  \bibfield  {author} {\bibinfo {author} {\bibfnamefont {J.}~\bibnamefont
  {Viljas}}\ and\ \bibinfo {author} {\bibfnamefont {T.}~\bibnamefont
  {Heikkil{\"a}}},\ }\href@noop {} {\bibfield  {journal} {\bibinfo  {journal}
  {Phys. Rev. B}\ }\textbf {\bibinfo {volume} {81}},\ \bibinfo {pages} {245404}
  (\bibinfo {year} {2010})}\BibitemShut {NoStop}%
\bibitem [{\citenamefont {Li}\ \emph {et~al.}(2010)\citenamefont {Li},
  \citenamefont {Barry}, \citenamefont {Zavada}, \citenamefont {Nardelli},\
  and\ \citenamefont {Kim}}]{XLi:APL10}%
  \BibitemOpen
  \bibfield  {author} {\bibinfo {author} {\bibfnamefont {X.}~\bibnamefont
  {Li}}, \bibinfo {author} {\bibfnamefont {E.}~\bibnamefont {Barry}}, \bibinfo
  {author} {\bibfnamefont {J.}~\bibnamefont {Zavada}}, \bibinfo {author}
  {\bibfnamefont {M.}~\bibnamefont {Nardelli}}, \ and\ \bibinfo {author}
  {\bibfnamefont {K.}~\bibnamefont {Kim}},\ }\href@noop {} {\bibfield
  {journal} {\bibinfo  {journal} {Appl. Phys. Lett.}\ }\textbf {\bibinfo
  {volume} {97}},\ \bibinfo {pages} {232105} (\bibinfo {year}
  {2010})}\BibitemShut {NoStop}%
\bibitem [{\citenamefont {Adam}\ \emph {et~al.}(2009)\citenamefont {Adam},
  \citenamefont {Hwang}, \citenamefont {Rossi},\ and\ \citenamefont
  {Das~Sarma}}]{SAdam:SSC09}%
  \BibitemOpen
  \bibfield  {author} {\bibinfo {author} {\bibfnamefont {S.}~\bibnamefont
  {Adam}}, \bibinfo {author} {\bibfnamefont {E.}~\bibnamefont {Hwang}},
  \bibinfo {author} {\bibfnamefont {E.}~\bibnamefont {Rossi}}, \ and\ \bibinfo
  {author} {\bibfnamefont {S.}~\bibnamefont {Das~Sarma}},\ }\href@noop {}
  {\bibfield  {journal} {\bibinfo  {journal} {Solid State Commun.}\ }\textbf
  {\bibinfo {volume} {149}},\ \bibinfo {pages} {1072} (\bibinfo {year}
  {2009})}\BibitemShut {NoStop}%
\bibitem [{\citenamefont {Adam}\ \emph {et~al.}(2007)\citenamefont {Adam},
  \citenamefont {Hwang}, \citenamefont {Galitski},\ and\ \citenamefont
  {Das~Sarma}}]{SAdam:PNAS07}%
  \BibitemOpen
  \bibfield  {author} {\bibinfo {author} {\bibfnamefont {S.}~\bibnamefont
  {Adam}}, \bibinfo {author} {\bibfnamefont {E.~H.}\ \bibnamefont {Hwang}},
  \bibinfo {author} {\bibfnamefont {V.~M.}\ \bibnamefont {Galitski}}, \ and\
  \bibinfo {author} {\bibfnamefont {S.}~\bibnamefont {Das~Sarma}},\ }\href@noop
  {} {\bibfield  {journal} {\bibinfo  {journal} {Proc. Natl. Acad. Sci.}\
  }\textbf {\bibinfo {volume} {104}},\ \bibinfo {pages} {18392} (\bibinfo
  {year} {2007})}\BibitemShut {NoStop}%
\bibitem [{\citenamefont {Lazzeri}\ and\ \citenamefont
  {Mauri}(2006)}]{MLazzeri:PRL06}%
  \BibitemOpen
  \bibfield  {author} {\bibinfo {author} {\bibfnamefont {M.}~\bibnamefont
  {Lazzeri}}\ and\ \bibinfo {author} {\bibfnamefont {F.}~\bibnamefont
  {Mauri}},\ }\href@noop {} {\bibfield  {journal} {\bibinfo  {journal} {Phys.
  Rev. Lett.}\ }\textbf {\bibinfo {volume} {97}},\ \bibinfo {pages} {266407}
  (\bibinfo {year} {2006})}\BibitemShut {NoStop}%
\bibitem [{\citenamefont {Pisana}\ \emph {et~al.}(2007)\citenamefont {Pisana},
  \citenamefont {Lazzeri}, \citenamefont {Casiraghi}, \citenamefont
  {Novoselov}, \citenamefont {Geim}, \citenamefont {Ferrari},\ and\
  \citenamefont {Mauri}}]{SPisana:NatureMaterials07}%
  \BibitemOpen
  \bibfield  {author} {\bibinfo {author} {\bibfnamefont {S.}~\bibnamefont
  {Pisana}}, \bibinfo {author} {\bibfnamefont {M.}~\bibnamefont {Lazzeri}},
  \bibinfo {author} {\bibfnamefont {C.}~\bibnamefont {Casiraghi}}, \bibinfo
  {author} {\bibfnamefont {K.}~\bibnamefont {Novoselov}}, \bibinfo {author}
  {\bibfnamefont {A.}~\bibnamefont {Geim}}, \bibinfo {author} {\bibfnamefont
  {A.}~\bibnamefont {Ferrari}}, \ and\ \bibinfo {author} {\bibfnamefont
  {F.}~\bibnamefont {Mauri}},\ }\href@noop {} {\bibfield  {journal} {\bibinfo
  {journal} {Nature Materials}\ }\textbf {\bibinfo {volume} {6}},\ \bibinfo
  {pages} {198} (\bibinfo {year} {2007})}\BibitemShut {NoStop}%
\bibitem [{\citenamefont {Ridley}(1999)}]{BKRidley:Book99}%
  \BibitemOpen
  \bibfield  {author} {\bibinfo {author} {\bibfnamefont {B.}~\bibnamefont
  {Ridley}},\ }\href@noop {} {\emph {\bibinfo {title} {Quantum processes in
  semiconductors}}}\ (\bibinfo  {publisher} {Oxford University Press, USA},\
  \bibinfo {year} {1999})\BibitemShut {NoStop}%
\bibitem [{\citenamefont {Perebeinos}\ and\ \citenamefont
  {Avouris}(2010)}]{VPerebeinos:PRB10}%
  \BibitemOpen
  \bibfield  {author} {\bibinfo {author} {\bibfnamefont {V.}~\bibnamefont
  {Perebeinos}}\ and\ \bibinfo {author} {\bibfnamefont {P.}~\bibnamefont
  {Avouris}},\ }\href@noop {} {\bibfield  {journal} {\bibinfo  {journal} {Phys.
  Rev. B}\ }\textbf {\bibinfo {volume} {81}},\ \bibinfo {pages} {195442}
  (\bibinfo {year} {2010})}\BibitemShut {NoStop}%
\bibitem [{\citenamefont {Liu}\ \emph {et~al.}(2008)\citenamefont {Liu},
  \citenamefont {Willis}, \citenamefont {Emtsev},\ and\ \citenamefont
  {Seyller}}]{YLiu:PRB08}%
  \BibitemOpen
  \bibfield  {author} {\bibinfo {author} {\bibfnamefont {Y.}~\bibnamefont
  {Liu}}, \bibinfo {author} {\bibfnamefont {R.~F.}\ \bibnamefont {Willis}},
  \bibinfo {author} {\bibfnamefont {K.~V.}\ \bibnamefont {Emtsev}}, \ and\
  \bibinfo {author} {\bibfnamefont {T.}~\bibnamefont {Seyller}},\ }\href@noop
  {} {\bibfield  {journal} {\bibinfo  {journal} {Phys. Rev. B}\ }\textbf
  {\bibinfo {volume} {78}},\ \bibinfo {pages} {201403} (\bibinfo {year}
  {2008})}\BibitemShut {NoStop}%
\bibitem [{\citenamefont {Liu}\ and\ \citenamefont
  {Willis}(2010)}]{YLiu:PRB10}%
  \BibitemOpen
  \bibfield  {author} {\bibinfo {author} {\bibfnamefont {Y.}~\bibnamefont
  {Liu}}\ and\ \bibinfo {author} {\bibfnamefont {R.~F.}\ \bibnamefont
  {Willis}},\ }\href@noop {} {\bibfield  {journal} {\bibinfo  {journal} {Phys.
  Rev. B}\ }\textbf {\bibinfo {volume} {81}},\ \bibinfo {pages} {081406}
  (\bibinfo {year} {2010})}\BibitemShut {NoStop}%
\bibitem [{\citenamefont {Fei}\ \emph {et~al.}(2011)\citenamefont {Fei},
  \citenamefont {Andreev}, \citenamefont {Bao}, \citenamefont {Zhang},
  \citenamefont {S.~McLeod}, \citenamefont {Wang}, \citenamefont {Stewart},
  \citenamefont {Zhao}, \citenamefont {Dominguez}, \citenamefont {Thiemens}
  \emph {et~al.}}]{ZFei:NL11}%
  \BibitemOpen
  \bibfield  {author} {\bibinfo {author} {\bibfnamefont {Z.}~\bibnamefont
  {Fei}}, \bibinfo {author} {\bibfnamefont {G.}~\bibnamefont {Andreev}},
  \bibinfo {author} {\bibfnamefont {W.}~\bibnamefont {Bao}}, \bibinfo {author}
  {\bibfnamefont {L.}~\bibnamefont {Zhang}}, \bibinfo {author} {\bibfnamefont
  {A.}~\bibnamefont {S.~McLeod}}, \bibinfo {author} {\bibfnamefont
  {C.}~\bibnamefont {Wang}}, \bibinfo {author} {\bibfnamefont {M.}~\bibnamefont
  {Stewart}}, \bibinfo {author} {\bibfnamefont {Z.}~\bibnamefont {Zhao}},
  \bibinfo {author} {\bibfnamefont {G.}~\bibnamefont {Dominguez}}, \bibinfo
  {author} {\bibfnamefont {M.}~\bibnamefont {Thiemens}},  \emph {et~al.},\
  }\href@noop {} {\bibfield  {journal} {\bibinfo  {journal} {Nano Lett.}\ ,\
  \bibinfo {pages} {4701}} (\bibinfo {year} {2011})}\BibitemShut {NoStop}%
\bibitem [{\citenamefont {Koch}\ \emph {et~al.}(2010)\citenamefont {Koch},
  \citenamefont {Seyller},\ and\ \citenamefont {Schaefer}}]{RJKoch:PRB10}%
  \BibitemOpen
  \bibfield  {author} {\bibinfo {author} {\bibfnamefont {R.~J.}\ \bibnamefont
  {Koch}}, \bibinfo {author} {\bibfnamefont {T.}~\bibnamefont {Seyller}}, \
  and\ \bibinfo {author} {\bibfnamefont {J.~A.}\ \bibnamefont {Schaefer}},\
  }\href@noop {} {\bibfield  {journal} {\bibinfo  {journal} {Phys. Rev. B}\
  }\textbf {\bibinfo {volume} {82}},\ \bibinfo {pages} {201413} (\bibinfo
  {year} {2010})}\BibitemShut {NoStop}%
\bibitem [{\citenamefont {Jackson}(1999)}]{JDJackson:Book99}%
  \BibitemOpen
  \bibfield  {author} {\bibinfo {author} {\bibfnamefont {J.~D.}\ \bibnamefont
  {Jackson}},\ }\href@noop {} {\emph {\bibinfo {title} {Classical
  Electrodynamics}}}\ (\bibinfo  {publisher} {New York: John Wiley \& Sons,
  Inc},\ \bibinfo {year} {1999})\BibitemShut {NoStop}%
\bibitem [{\citenamefont {Kim}\ \emph {et~al.}(1978)\citenamefont {Kim},
  \citenamefont {Das},\ and\ \citenamefont {Senturia}}]{MEKim:PRB78}%
  \BibitemOpen
  \bibfield  {author} {\bibinfo {author} {\bibfnamefont {M.~E.}\ \bibnamefont
  {Kim}}, \bibinfo {author} {\bibfnamefont {A.}~\bibnamefont {Das}}, \ and\
  \bibinfo {author} {\bibfnamefont {S.~D.}\ \bibnamefont {Senturia}},\
  }\href@noop {} {\bibfield  {journal} {\bibinfo  {journal} {Phys. Rev. B}\
  }\textbf {\bibinfo {volume} {18}},\ \bibinfo {pages} {6890} (\bibinfo {year}
  {1978})}\BibitemShut {NoStop}%
\bibitem [{\citenamefont {Hwang}\ and\ \citenamefont
  {Das~Sarma}(2007)}]{EHHwang:PRB07}%
  \BibitemOpen
  \bibfield  {author} {\bibinfo {author} {\bibfnamefont {E.~H.}\ \bibnamefont
  {Hwang}}\ and\ \bibinfo {author} {\bibfnamefont {S.}~\bibnamefont
  {Das~Sarma}},\ }\href@noop {} {\bibfield  {journal} {\bibinfo  {journal}
  {Phys. Rev. B}\ }\textbf {\bibinfo {volume} {75}},\ \bibinfo {pages} {205418}
  (\bibinfo {year} {2007})}\BibitemShut {NoStop}%
\bibitem [{\citenamefont {Wunsch}\ \emph {et~al.}(2006)\citenamefont {Wunsch},
  \citenamefont {Stauber}, \citenamefont {Sols},\ and\ \citenamefont
  {Guinea}}]{BWunsch:NJP06}%
  \BibitemOpen
  \bibfield  {author} {\bibinfo {author} {\bibfnamefont {B.}~\bibnamefont
  {Wunsch}}, \bibinfo {author} {\bibfnamefont {T.}~\bibnamefont {Stauber}},
  \bibinfo {author} {\bibfnamefont {F.}~\bibnamefont {Sols}}, \ and\ \bibinfo
  {author} {\bibfnamefont {F.}~\bibnamefont {Guinea}},\ }\href@noop {}
  {\bibfield  {journal} {\bibinfo  {journal} {New J. Phys.}\ }\textbf {\bibinfo
  {volume} {8}},\ \bibinfo {pages} {318} (\bibinfo {year} {2006})}\BibitemShut
  {NoStop}%
\bibitem [{\citenamefont {Zou}\ \emph {et~al.}(2010)\citenamefont {Zou},
  \citenamefont {Hong}, \citenamefont {Keefer},\ and\ \citenamefont
  {Zhu}}]{KZou:PRL10}%
  \BibitemOpen
  \bibfield  {author} {\bibinfo {author} {\bibfnamefont {K.}~\bibnamefont
  {Zou}}, \bibinfo {author} {\bibfnamefont {X.}~\bibnamefont {Hong}}, \bibinfo
  {author} {\bibfnamefont {D.}~\bibnamefont {Keefer}}, \ and\ \bibinfo {author}
  {\bibfnamefont {J.}~\bibnamefont {Zhu}},\ }\href@noop {} {\bibfield
  {journal} {\bibinfo  {journal} {Phys. Rev. Lett.}\ }\textbf {\bibinfo
  {volume} {105}},\ \bibinfo {pages} {126601} (\bibinfo {year}
  {2010})}\BibitemShut {NoStop}%
\bibitem [{\citenamefont {Hwang}\ \emph {et~al.}(2010)\citenamefont {Hwang},
  \citenamefont {Sensarma},\ and\ \citenamefont {Das~Sarma}}]{EHHwang:PRB10}%
  \BibitemOpen
  \bibfield  {author} {\bibinfo {author} {\bibfnamefont {E.~H.}\ \bibnamefont
  {Hwang}}, \bibinfo {author} {\bibfnamefont {R.}~\bibnamefont {Sensarma}}, \
  and\ \bibinfo {author} {\bibfnamefont {S.}~\bibnamefont {Das~Sarma}},\
  }\href@noop {} {\bibfield  {journal} {\bibinfo  {journal} {Phys. Rev. B}\
  }\textbf {\bibinfo {volume} {82}},\ \bibinfo {pages} {195406} (\bibinfo
  {year} {2010})}\BibitemShut {NoStop}%
\bibitem [{\citenamefont {Shishir}\ and\ \citenamefont
  {Ferry}(2009)}]{RSShishir:JPhys09}%
  \BibitemOpen
  \bibfield  {author} {\bibinfo {author} {\bibfnamefont {R.}~\bibnamefont
  {Shishir}}\ and\ \bibinfo {author} {\bibfnamefont {D.}~\bibnamefont
  {Ferry}},\ }\href@noop {} {\bibfield  {journal} {\bibinfo  {journal} {J.
  Phys. C}\ }\textbf {\bibinfo {volume} {21}},\ \bibinfo {pages} {232204}
  (\bibinfo {year} {2009})}\BibitemShut {NoStop}%
\bibitem [{\citenamefont {Tan}\ \emph {et~al.}(2007)\citenamefont {Tan},
  \citenamefont {Zhang}, \citenamefont {Bolotin}, \citenamefont {Zhao},
  \citenamefont {Adam}, \citenamefont {Hwang}, \citenamefont {Das~Sarma},
  \citenamefont {Stormer},\ and\ \citenamefont {Kim}}]{YWTan:PRL07}%
  \BibitemOpen
  \bibfield  {author} {\bibinfo {author} {\bibfnamefont {Y.}~\bibnamefont
  {Tan}}, \bibinfo {author} {\bibfnamefont {Y.}~\bibnamefont {Zhang}}, \bibinfo
  {author} {\bibfnamefont {K.}~\bibnamefont {Bolotin}}, \bibinfo {author}
  {\bibfnamefont {Y.}~\bibnamefont {Zhao}}, \bibinfo {author} {\bibfnamefont
  {S.}~\bibnamefont {Adam}}, \bibinfo {author} {\bibfnamefont {E.}~\bibnamefont
  {Hwang}}, \bibinfo {author} {\bibfnamefont {S.}~\bibnamefont {Das~Sarma}},
  \bibinfo {author} {\bibfnamefont {H.}~\bibnamefont {Stormer}}, \ and\
  \bibinfo {author} {\bibfnamefont {P.}~\bibnamefont {Kim}},\ }\href@noop {}
  {\bibfield  {journal} {\bibinfo  {journal} {Phys. Rev. Lett.}\ }\textbf
  {\bibinfo {volume} {99}},\ \bibinfo {pages} {246803} (\bibinfo {year}
  {2007})}\BibitemShut {NoStop}%
\bibitem [{\citenamefont {Jang}\ \emph {et~al.}(2008)\citenamefont {Jang},
  \citenamefont {Adam}, \citenamefont {Chen}, \citenamefont {Williams},
  \citenamefont {Das~Sarma},\ and\ \citenamefont {Fuhrer}}]{CJang:PRL08}%
  \BibitemOpen
  \bibfield  {author} {\bibinfo {author} {\bibfnamefont {C.}~\bibnamefont
  {Jang}}, \bibinfo {author} {\bibfnamefont {S.}~\bibnamefont {Adam}}, \bibinfo
  {author} {\bibfnamefont {J.}~\bibnamefont {Chen}}, \bibinfo {author}
  {\bibfnamefont {E.}~\bibnamefont {Williams}}, \bibinfo {author}
  {\bibfnamefont {S.}~\bibnamefont {Das~Sarma}}, \ and\ \bibinfo {author}
  {\bibfnamefont {M.}~\bibnamefont {Fuhrer}},\ }\href@noop {} {\bibfield
  {journal} {\bibinfo  {journal} {Phys. Rev. Lett.}\ }\textbf {\bibinfo
  {volume} {101}},\ \bibinfo {pages} {146805} (\bibinfo {year}
  {2008})}\BibitemShut {NoStop}%
\bibitem [{\citenamefont {Ponomarenko}\ \emph {et~al.}(2009)\citenamefont
  {Ponomarenko}, \citenamefont {Yang}, \citenamefont {Mohiuddin}, \citenamefont
  {Katsnelson}, \citenamefont {Novoselov}, \citenamefont {Morozov},
  \citenamefont {Zhukov}, \citenamefont {Schedin}, \citenamefont {Hill},\ and\
  \citenamefont {Geim}}]{LAPonomarenko:PRL09}%
  \BibitemOpen
  \bibfield  {author} {\bibinfo {author} {\bibfnamefont {L.}~\bibnamefont
  {Ponomarenko}}, \bibinfo {author} {\bibfnamefont {R.}~\bibnamefont {Yang}},
  \bibinfo {author} {\bibfnamefont {T.}~\bibnamefont {Mohiuddin}}, \bibinfo
  {author} {\bibfnamefont {M.}~\bibnamefont {Katsnelson}}, \bibinfo {author}
  {\bibfnamefont {K.}~\bibnamefont {Novoselov}}, \bibinfo {author}
  {\bibfnamefont {S.}~\bibnamefont {Morozov}}, \bibinfo {author} {\bibfnamefont
  {A.}~\bibnamefont {Zhukov}}, \bibinfo {author} {\bibfnamefont
  {F.}~\bibnamefont {Schedin}}, \bibinfo {author} {\bibfnamefont
  {E.}~\bibnamefont {Hill}}, \ and\ \bibinfo {author} {\bibfnamefont
  {A.}~\bibnamefont {Geim}},\ }\href@noop {} {\bibfield  {journal} {\bibinfo
  {journal} {Phys. Rev. Lett.}\ }\textbf {\bibinfo {volume} {102}},\ \bibinfo
  {pages} {206603} (\bibinfo {year} {2009})}\BibitemShut {NoStop}%
\bibitem [{\citenamefont {Garces}\ \emph {et~al.}(2011)\citenamefont {Garces},
  \citenamefont {Wheeler}, \citenamefont {Hite}, \citenamefont {Jernigan},
  \citenamefont {Tedesco}, \citenamefont {Nepal}, \citenamefont {Eddy},\ and\
  \citenamefont {Gaskill}}]{NYGarces:JAP11}%
  \BibitemOpen
  \bibfield  {author} {\bibinfo {author} {\bibfnamefont {N.}~\bibnamefont
  {Garces}}, \bibinfo {author} {\bibfnamefont {V.}~\bibnamefont {Wheeler}},
  \bibinfo {author} {\bibfnamefont {J.}~\bibnamefont {Hite}}, \bibinfo {author}
  {\bibfnamefont {G.}~\bibnamefont {Jernigan}}, \bibinfo {author}
  {\bibfnamefont {J.}~\bibnamefont {Tedesco}}, \bibinfo {author} {\bibfnamefont
  {N.}~\bibnamefont {Nepal}}, \bibinfo {author} {\bibfnamefont
  {C.}~\bibnamefont {Eddy}}, \ and\ \bibinfo {author} {\bibfnamefont
  {D.}~\bibnamefont {Gaskill}},\ }\href@noop {} {\bibfield  {journal} {\bibinfo
   {journal} {J. Appl. Phys.}\ }\textbf {\bibinfo {volume} {109}},\ \bibinfo
  {pages} {124304} (\bibinfo {year} {2011})}\BibitemShut {NoStop}%
\bibitem [{\citenamefont {Gilat}\ and\ \citenamefont
  {Raubenheimer}(1966)}]{GGilat:PR66}%
  \BibitemOpen
  \bibfield  {author} {\bibinfo {author} {\bibfnamefont {G.}~\bibnamefont
  {Gilat}}\ and\ \bibinfo {author} {\bibfnamefont {L.}~\bibnamefont
  {Raubenheimer}},\ }\href@noop {} {\bibfield  {journal} {\bibinfo  {journal}
  {Phys. Rev.}\ }\textbf {\bibinfo {volume} {144}},\ \bibinfo {pages} {390}
  (\bibinfo {year} {1966})}\BibitemShut {NoStop}%
\bibitem [{\citenamefont {Fallahazad}\ \emph {et~al.}(2010)\citenamefont
  {Fallahazad}, \citenamefont {Kim}, \citenamefont {Colombo},\ and\
  \citenamefont {Tutuc}}]{BFallahazad:APL10}%
  \BibitemOpen
  \bibfield  {author} {\bibinfo {author} {\bibfnamefont {B.}~\bibnamefont
  {Fallahazad}}, \bibinfo {author} {\bibfnamefont {S.}~\bibnamefont {Kim}},
  \bibinfo {author} {\bibfnamefont {L.}~\bibnamefont {Colombo}}, \ and\
  \bibinfo {author} {\bibfnamefont {E.}~\bibnamefont {Tutuc}},\ }\href@noop {}
  {\bibfield  {journal} {\bibinfo  {journal} {Appl. Phys. Lett.}\ }\textbf
  {\bibinfo {volume} {97}},\ \bibinfo {pages} {123105} (\bibinfo {year}
  {2010})}\BibitemShut {NoStop}%
\bibitem [{\citenamefont {Jandhyala}\ \emph {et~al.}(2012)\citenamefont
  {Jandhyala}, \citenamefont {Mordi}, \citenamefont {Lee}, \citenamefont {Lee},
  \citenamefont {Floresca}, \citenamefont {Cha}, \citenamefont {Ahn},
  \citenamefont {Wallace}, \citenamefont {Chabal}, \citenamefont {Kim} \emph
  {et~al.}}]{SJandhyala:ACSNano12}%
  \BibitemOpen
  \bibfield  {author} {\bibinfo {author} {\bibfnamefont {S.}~\bibnamefont
  {Jandhyala}}, \bibinfo {author} {\bibfnamefont {G.}~\bibnamefont {Mordi}},
  \bibinfo {author} {\bibfnamefont {B.}~\bibnamefont {Lee}}, \bibinfo {author}
  {\bibfnamefont {G.}~\bibnamefont {Lee}}, \bibinfo {author} {\bibfnamefont
  {C.}~\bibnamefont {Floresca}}, \bibinfo {author} {\bibfnamefont
  {P.}~\bibnamefont {Cha}}, \bibinfo {author} {\bibfnamefont {J.}~\bibnamefont
  {Ahn}}, \bibinfo {author} {\bibfnamefont {R.}~\bibnamefont {Wallace}},
  \bibinfo {author} {\bibfnamefont {Y.}~\bibnamefont {Chabal}}, \bibinfo
  {author} {\bibfnamefont {M.}~\bibnamefont {Kim}},  \emph {et~al.},\
  }\href@noop {} {\bibfield  {journal} {\bibinfo  {journal} {ACS Nano}\ }
  (\bibinfo {year} {2012})}\BibitemShut {NoStop}%
\bibitem [{\citenamefont {Ren}\ \emph {et~al.}(2003)\citenamefont {Ren},
  \citenamefont {Fischetti}, \citenamefont {Gusev}, \citenamefont {Cartier},\
  and\ \citenamefont {Chudzik}}]{ZRen:IEDM03}%
  \BibitemOpen
  \bibfield  {author} {\bibinfo {author} {\bibfnamefont {Z.}~\bibnamefont
  {Ren}}, \bibinfo {author} {\bibfnamefont {M.}~\bibnamefont {Fischetti}},
  \bibinfo {author} {\bibfnamefont {E.}~\bibnamefont {Gusev}}, \bibinfo
  {author} {\bibfnamefont {E.}~\bibnamefont {Cartier}}, \ and\ \bibinfo
  {author} {\bibfnamefont {M.}~\bibnamefont {Chudzik}},\ }in\ \href@noop {}
  {\emph {\bibinfo {booktitle} {Electron Devices Meeting, 2003. IEDM'03
  Technical Digest. IEEE International}}}\ (\bibinfo {organization} {IEEE},\
  \bibinfo {year} {2003})\ pp.\ \bibinfo {pages} {33--2}\BibitemShut {NoStop}%
\bibitem [{\citenamefont {Zhu}\ \emph {et~al.}(2009)\citenamefont {Zhu},
  \citenamefont {Perebeinos}, \citenamefont {Freitag},\ and\ \citenamefont
  {Avouris}}]{WZhu:PRB09}%
  \BibitemOpen
  \bibfield  {author} {\bibinfo {author} {\bibfnamefont {W.}~\bibnamefont
  {Zhu}}, \bibinfo {author} {\bibfnamefont {V.}~\bibnamefont {Perebeinos}},
  \bibinfo {author} {\bibfnamefont {M.}~\bibnamefont {Freitag}}, \ and\
  \bibinfo {author} {\bibfnamefont {P.}~\bibnamefont {Avouris}},\ }\href@noop
  {} {\bibfield  {journal} {\bibinfo  {journal} {Phys. Rev. B}\ }\textbf
  {\bibinfo {volume} {80}},\ \bibinfo {pages} {235402} (\bibinfo {year}
  {2009})}\BibitemShut {NoStop}%
\bibitem [{\citenamefont {Li}\ \emph {et~al.}(2009)\citenamefont {Li},
  \citenamefont {Cai}, \citenamefont {An}, \citenamefont {Kim}, \citenamefont
  {Nah}, \citenamefont {Yang}, \citenamefont {Piner}, \citenamefont
  {Velamakanni}, \citenamefont {Jung}, \citenamefont {Tutuc} \emph
  {et~al.}}]{XLi:Science09}%
  \BibitemOpen
  \bibfield  {author} {\bibinfo {author} {\bibfnamefont {X.}~\bibnamefont
  {Li}}, \bibinfo {author} {\bibfnamefont {W.}~\bibnamefont {Cai}}, \bibinfo
  {author} {\bibfnamefont {J.}~\bibnamefont {An}}, \bibinfo {author}
  {\bibfnamefont {S.}~\bibnamefont {Kim}}, \bibinfo {author} {\bibfnamefont
  {J.}~\bibnamefont {Nah}}, \bibinfo {author} {\bibfnamefont {D.}~\bibnamefont
  {Yang}}, \bibinfo {author} {\bibfnamefont {R.}~\bibnamefont {Piner}},
  \bibinfo {author} {\bibfnamefont {A.}~\bibnamefont {Velamakanni}}, \bibinfo
  {author} {\bibfnamefont {I.}~\bibnamefont {Jung}}, \bibinfo {author}
  {\bibfnamefont {E.}~\bibnamefont {Tutuc}},  \emph {et~al.},\ }\href@noop {}
  {\bibfield  {journal} {\bibinfo  {journal} {Science}\ }\textbf {\bibinfo
  {volume} {324}},\ \bibinfo {pages} {1312} (\bibinfo {year}
  {2009})}\BibitemShut {NoStop}%
\bibitem [{\citenamefont {Krishan}\ and\ \citenamefont
  {Ritchie}(1970)}]{VKrishan:PRL70}%
  \BibitemOpen
  \bibfield  {author} {\bibinfo {author} {\bibfnamefont {V.}~\bibnamefont
  {Krishan}}\ and\ \bibinfo {author} {\bibfnamefont {R.~H.}\ \bibnamefont
  {Ritchie}},\ }\href@noop {} {\bibfield  {journal} {\bibinfo  {journal} {Phys.
  Rev. Lett.}\ }\textbf {\bibinfo {volume} {24}},\ \bibinfo {pages} {1117}
  (\bibinfo {year} {1970})}\BibitemShut {NoStop}%
\bibitem [{\citenamefont {Adam}\ \emph {et~al.}(2011)\citenamefont {Adam},
  \citenamefont {Jung}, \citenamefont {Klimov}, \citenamefont {Zhitenev},
  \citenamefont {Stroscio},\ and\ \citenamefont {Stiles}}]{SAdam:PRB11}%
  \BibitemOpen
  \bibfield  {author} {\bibinfo {author} {\bibfnamefont {S.}~\bibnamefont
  {Adam}}, \bibinfo {author} {\bibfnamefont {S.}~\bibnamefont {Jung}}, \bibinfo
  {author} {\bibfnamefont {N.}~\bibnamefont {Klimov}}, \bibinfo {author}
  {\bibfnamefont {N.}~\bibnamefont {Zhitenev}}, \bibinfo {author}
  {\bibfnamefont {J.}~\bibnamefont {Stroscio}}, \ and\ \bibinfo {author}
  {\bibfnamefont {M.}~\bibnamefont {Stiles}},\ }\href@noop {} {\bibfield
  {journal} {\bibinfo  {journal} {Phys. Rev. B}\ }\textbf {\bibinfo {volume}
  {84}},\ \bibinfo {pages} {235421} (\bibinfo {year} {2011})}\BibitemShut
  {NoStop}%
\end{thebibliography}%

\end{document}